\documentclass[aps,pre,amsmath,superscriptaddress,amssymb,twocolumn,showpacs,showkeys,longbibliography]{revtex4-1}  
\usepackage{color}
\usepackage{enumitem}
\usepackage{graphicx,subfigure}
\usepackage[normalem]{ulem}
\usepackage{lineno}
\usepackage{booktabs}
   \usepackage{array}
    \usepackage{multirow}
    \usepackage{hhline}
    \usepackage{makecell}
     \usepackage[symbol]{footmisc}

\newif\ifHighlitedChanges
\def\ifHighlitedChanges{\iftrue}
\ifHighlitedChanges
   
  \def\STRIKE#1{{\color{blue}\sout{#1}}}
\else
  
  \def\STRIKE#1{\relax}
\fi

\newif\ifHighlitedChanges
\def\ifHighlitedChanges{\iftrue}
\ifHighlitedChanges

\else

\fi

\begin{document}
\title{Anomalous Vapor and Ice Nucleation in Water at Negative Pressures: A Classical Density Functional Theory Study}
\author{Yuvraj Singh}
\affiliation{Department of Physics, Indian Institute of Science Education and Research (IISER) Tirupati, Tirupati, Andhra Pradesh 517507, India}
\author{Mantu Santra}
\affiliation{School of Chemical and Materials Sciences, Indian Institute of Technology Goa, Goa 403401, India}
\author{Rakesh S. Singh}
\email{rssingh@iisertirupati.ac.in}
\affiliation{Department of Chemistry, Indian Institute of Science Education and Research (IISER) Tirupati, Tirupati, Andhra Pradesh 517507, India}
\date{\today}

\begin{abstract}
Water is well-known for its anomalous thermodynamic behavior at both normal and extreme (such as supercooled and negative pressure) conditions. However, in contrast to the abundance of work on the anomalous behavior of water, the relationship between the water's thermodynamic anomalies and kinetics of phase transition from metastable (supercooled and/or negative pressure) water is relatively unexplored. In this work, we have employed classical density functional theory to provide a unified and coherent picture of nucleation (both vapor and ice) from metastable water, especially at negative pressure conditions. Our results suggest a peculiar non-monotonic temperature dependence of liquid-vapor surface tension at temperatures where liquid-vapor coexistence is metastable with respect to the ice phase. The vapor nucleation barrier on isochoric cooling also shows a non-monotonic temperature dependence. We further note that, for low density isochores, the temperature of minimum vapor nucleation barrier ($T_{\Delta \Omega_{\rm v/min}^*}$) does not coincide with the temperature of maximum density (TMD) where metastability is maximum. The temperature difference between the $T_{\Delta \Omega_{\rm v/min}^*}$ and the TMD, however, decreases with increasing the density of the isochore, suggesting a strong correlation between the propensity of cavitation and metastability of the liquid water at high densities. The vapor nucleation barrier along isobaric cooling shows an interesting crossover behavior where it first increases on lowering the temperature and then shows a non-monotonic behavior in the vicinity of the Widom line on further lowering the temperature. Our results on the ice nucleation from metastable water show an anomalous retracing behavior of the ice nucleation barrier along isotherms and theoretically validate the recent findings that the reentrant ice(Ih)-liquid coexistence can induce a drastic change in the kinetics of ice nucleation. Thus, this study establishes a direct connection between the water's thermodynamic anomalies and the (vapor and ice) nucleation kinetics. In addition, this study also provides deeper insights into the origin of the isothermal compressibility maximum on isochoric cooling.                                                                                                                                                                                                                                                                                                                                                                                                                                                                                                                                                                                                                                                                    
\end{abstract}

\maketitle

\section{Introduction}\label{intro}
Freezing (ice formation) and boiling of liquid water are unarguably the most ubiquitous phase transitions in nature~\cite{pablo_book, pablo_nat_2006}. The ice formations are relevant to a wide range of disciplines ranging from materials to biological to food and climate \& earth sciences~\cite{wilson2010type, wilson2010type, gettelman2010global, broekaert2011seafood, bintanja2013important, morris2013controlled, bar2016ice, amir_ice_rev_2020, nitzbon2020fast}. Recently, phase transitions in water at extreme conditions, such as cavitation transition at negative pressure ($P$) conditions, have drawn considerable interest~\cite{angell_sci, pablo_nature_2013, caupin2005liquid, caupin2006cavitation, frederic_natphys_2013, frederic_pnas_2016}. The cavitation transitions are also relevant to many processes in nature, like water transport in natural and synthetic trees~\cite{negative_p_nature_2008, negative_p_prl_2012}, poration of cell membranes~\cite{cell_jpcb_2015}, and sonocrystallization of ice~\cite{sono_xtal_apl_1998, sono_xtal_us_2012}. The cavitation transitions are also used in experiments to locate the temperature of the maximum density (TMD) line at negative pressure conditions~\cite{angell_sci, frederic_natphys_2013}. The shape of the TMD line at negative pressures has the potential to provide important insights into the origin of water's thermodynamic anomalies and distinguish different competing scenarios proposed to explain the origin of water's anomalies at supercooled conditions~\cite{pablo_book, bagchi_book, pablo_rev_2003, angell_1973, angell_1982, tombari_anomaly_1999, kanno_angell_1979}.   

The pioneering work of Angell and coworkers~\cite{angell_sci} on water trapped in mineral inclusions suggests that water can sustain strong negative pressure up to $-140$ MPa before it breaks by cavitation. This ability to withstand strong negative pressure is attributed to the strong cohesive forces between the molecules in the liquid. Recent computational and experimental studies further suggest rich anomalous thermodynamic behavior of liquid water at negative pressure conditions, including doubly metastable states where liquid water is simultaneously metastable with respect to both vapor and ice phases. The thermodynamic anomalies at negative pressures include the retracing behavior of the TMD line, the anomalous increase of the thermodynamic response functions (such as isothermal compressibility $\kappa_T$, and isobaric heat capacity $C_P$) on isobaric cooling, and the sound velocity minimum and isothermal compressibility maximum on isochoric cooling~\cite{stanley_jcp_2001, frederic_pnas_2014, caupin_rev_2015, singh_tip4p_2017, singh_tip4p_2017_2, frederic_jpcl_2017}. The anomalous increase of the thermodynamic response functions on isobaric cooling at ambient and negative pressures is often attributed to the existence of a hypothetical liquid-liquid transition (LLT) at high pressures and the associated Widom line~\cite{poole_1992,pablo_nature_2014, palmer_chem_rev_2018, gallo_chem_rev_2016, singh_tip4p_2017, frederic_jpcl_2017}. The anomalous temperature ($T$) dependence of the sound velocity and isothermal compressibility along isochores at negative pressures was attributed recently to the peculiar reentrant behavior of the liquid-vapor (LV) spinodal in the temperature-density ($\rho-T$) plane~\cite{singh_tip4p_2017_2}. 
    
In contrast to the water's anomalous phase behvaior which is well-studied at ambient as well as extreme conditions in both the computer simulation and experimental studies~\cite{speedy_1976, poole_1992, caupin_jcp_2010, pablo_nature_2014, anisimov_mb, anisimov_st2, singh_tip4p_2016, singh_tip4p_2017, singh_tip4p_2017_2, frederic_pnas_2014, frederic_jpcl_2017, stanley_jcp_2001, negative_P_book, gallo_chem_rev_2016, palmer_chem_rev_2018, pablo_ml_water_2022}, very few studies have explored the nucleation (vapor and ice) from metastable liquid water near ambient~\cite{ohmine_2002, molinero_2011, vega_jcp_2012, vega_ice_jacs_2013, amir_pnas_2015, zaragoza_2015, amir_ice_pccp_2016, amir_ice_pnas_2017, vega_ice_jpcl_2017, dellago_ice_pccp_2018, michaelides_2019, palmer_2019, jeremy_ice_jcp_2022} and negative pressure conditions~\cite{angell_sci, frederic_natphys_2013, frederic_pnas_2016, abascal_jcp_2013, gonzalez_jcp_2014, gonzalez_jcp_2015} and even fewer have explored the interplay between the observed water's thermodynamic anomalies and the nucleation barrier from metastable liquid water~\cite{angell_sci, frederic_natphys_2013, vega_prl_2021, singh_2014, poole_2015}. In a recent seminal study, using computer simulations of TIP4P/Ice water, Vega and coworkers~\cite{vega_prl_2021} studied ice nucleation from metastable liquid water at pressures ranging from high to negative pressures close to the LV spinodal. They reported an anomalous (non-monotonic) dependence of the ice nucleation barrier on pressure along isotherms where the nucleation barrier reaches a minimum at negative pressures in the doubly metastable region. This study also predicted an anomalous reentrant behavior of the homogeneous ice nucleation line.    
 
 In the fluid inclusion in minerals experiments, Angell and coworkers~\cite{angell_sci} estimated the TMD for low-density isochores at negative pressures by measuring the temperature variation of the cavitation rate. The assumption was -- at the TMD, liquid water is maximally stretched and is expected to be more prone to cavitation (vapor nucleation). Thus, the temperature of the minimum barrier of nucleation on isochoric cooling would be the TMD. In subsequent work on a single low density isochore, Caupin and coworkers~\cite{frederic_natphys_2013} observed that the temperature of the minimum barrier does not coincide with the TMD suggesting that maximally stretched water is not necessarily more prone to cavitation. Thus, these studies suggest an interesting interplay between anomalous thermodynamic behavior (such as, TMD) and the (vapor) nucleation barrier. It is also worth noting that the classical nucleation theory (CNT)~\cite{pablo_book, becker-doring-1935, frenkel-book-1955, oxtoby_cnt_rev_1992} was used to estimate the nucleation barrier in the cavitation experiment by Caupin and coworker~\cite{frederic_natphys_2013}, and also in the computational study by Vega and coworkers~\cite{vega_prl_2021}. Although CNT provides a physically simple framework to estimate the nucleation barrier, its validity at deeply metastable conditions is questionable as the thermodynamic properties of the critical cluster may differ substantially from the stable bulk phase at these conditions~\cite{oxtoby_cnt_rev_1992, parrinello_prl_2005, pablo_nat_2006, suman_prl_2007, mantu_pre_2011, mantu_jstat_2011}. 
 
In this work, we have employed classical density functional theory (CDFT)~\cite{pablo_book, ch_1958, pablo_dft_2001} to provide a unified and coherent picture of nucleation (both vapor and ice) in metastable liquid water, especially at negative pressure conditions. The main aim of this work is to study the interplay between the water's rich thermodynamic anomalies (such as, isothermal compressibility maximum on isobaric and isochoric cooling, TMD, and reentrant behavior of ice-liquid coexistence line) and the free energy barrier of nucleation (vapor and ice) from the negative pressure (or, stretched) water. The computational studies of ice nucleation from the metastable water is turned out to be a challenging task due to the sluggish structural relaxation of water at low temperatures, and direct estimation of the nucleation barrier or rate often involves computationally intensive methods, like forward flux sampling~\cite{ffs} and umbrella sampling~\cite{us}. The experimental studies of the ice nucleation barrier and its dependence on the thermodynamic parameters ($P, T$) are restricted by the rapid ice crystallization from the metastable water. Thus, CDFT  provides a robust alternative to the computer simulation and experimental approaches as it is not restricted by the aforementioned limitations often encountered in experiments and computer simulation studies. The CDFT-based approach also enables one to check the validity of the CNT at highly supercooled or metastable conditions (close to the LV spinodal or the stability limit of liquid water) as it is not restricted by the assumptions involved in the CNT. Of course, the reliability of the CDFT predictions depends on the robustness of the representative free energy functional in describing the phase behavior of the system. Here, we have used the phenomenological microscopic model developed by Truskett et al.~\cite{pablo_jcp_1999} as this model accurately captures the (fluid) phase behavior of water near ambient conditions and also makes some interesting predictions under supercooled and negative pressure conditions, including the existence of the (hypothesized) liquid-liquid critical point (LLCP)~\cite{palmer_2019, poole_1992, pablo_nature_2014,singh_tip4p_2017_2}.                    

The rest of this paper is organized as follows. Section~\ref{model_phase} details the phenomenological microscopic model used for the CDFT-based calculations. The liquid-vapor surface tension at different temperatures ranging from normal to supercooled (coexisting liquid and vapor phases are metastable with respect to the ice phase) conditions along the (liquid-vapor) coexistence line are discussed in Section~\ref{lvst}. In Sections~~\ref{lvv} and~\ref{lvp}, we discuss the anomalous temperature dependence of the vapor nucleation barrier from stretched water on isochoric and isobaric cooling, respectively. The anomalous ice nucleation behavior from the metastable and doubly metastable water is discussed in Section~\ref{Is}. In Section~\ref{phase_diag}, we present a phase diagram summarizing the anomalous nucleation (vapor and ice) from metastable liquid water, and the major findings of this work are outlined in Section~\ref{conclusions}.             
\section{Model and Phase Behavior}\label{model_phase}
\subsection{Microscopic Model}
As discussed in the previous section, we have employed the microscopic phenomenological model developed by Truskett et al.~\cite{pablo_jcp_1999} to explore the interplay between the thermodynamic anomalies and nucleation barrier in stretched and doubly metastable water. Here, we provide some key realistic features of this model: (i) to acknowledge hydrogen bonds' (H-bonds') directionality which leads to the low-density environment in the close vicinity of H-bonds, the model is geometrically designed to have a cavity of radius $r_i$ ($1.01 \sigma$, where $\sigma = 3.11 \AA$), (ii) the cut-off distance $r_o$ ($1.04 \sigma$) for the H-bond formation between two molecules is assigned to address the short-range interaction of hydrogen molecules, (iii) as H-bonds are highly orientation-dependent, the model is designed to constrain the water molecules within an angle of $\phi^*$ at the line joining the centre of water molecules, and (iv) the presence of non-bonding molecules in the H-bonding shell crowds the central molecule and hence weakens the H-bonds. To incorporate this feature in the model, an energetic penalty $\epsilon_{\rm p}$ is assigned to each non-bonding molecule in the H-bonding shell. The stability of hydrogen is simplified as $-\epsilon_j = -\epsilon_{\rm max} + (j-1) \epsilon_{\rm p}$, where $\epsilon_{\rm max}$ is the maximum stability of the H-bond, and $j$ is the number of non-bonding molecules in the H-bonding shell of the central molecule.

The interaction potential between the molecules ($\Phi$) can be decomposed into three different contributions: (i) hard sphere interaction  ($\Phi_{\rm HS}$), (ii) dispersion interaction ($\Phi_{\rm DI}$), and (iii) H-bond interaction ($\Phi_{\rm HB}$). That is,  $\Phi =$ $\Phi_{\rm HS}$ + $\Phi_{\rm DI}$ + $\Phi_{\rm HB}$. Considering these contributions and geometric criteria of the model, and after approximating the hard sphere, dispersion, and H-bonding interactions, Truskett et al.~\cite{pablo_jcp_1999} derived the following expression for the canonical partition function, $Z$ (see Ref.~\cite{pablo_jcp_1999} for the details), 
\begin{equation} \label{eq:2}
Z(N,V,T) = \left(\frac{1}{N! \Lambda^{3N}}\right) V_{\rm ex}^N  \exp(N\beta\rho a) (4\pi)^N \prod_{j=1}^8{y_j^{Np_j}},
\end{equation}
where $N$ is the number of particles, $V$ is the volume of the system, $T$ is the absolute temperature, $\Lambda$ is the thermal wavelength, $\beta = 1/\emph{k}_{\rm B}T$ ($k_{\rm B}$ is Boltzmann's constant), $y_j = \left[1+ \frac{j}{4}(1-\cos \phi^*)^2 [\exp(\beta\epsilon_j) - 1] \right]$, and $V_{\rm ex} = V - Nb$ ($b$ is the excluded volume per particle). The values of the parameters are chosen as $\epsilon_{\rm max} = 23~\rm kJ/mol$, $\epsilon_{\rm p} = 3~\rm kJ/mol$, $a = 0.269~\rm Pa m^6 mol^{-2}$, and $\phi^* = 0.175~\rm rad $ (see Ref.~\cite{pablo_jcp_1999}). The Helmholtz free energy is $F_{\rm fl} (N,V,T) = -k_{\rm B}T \ln Z(N,V,T)$. 

\begin{figure} 
    \centering
          \includegraphics[width=\linewidth]{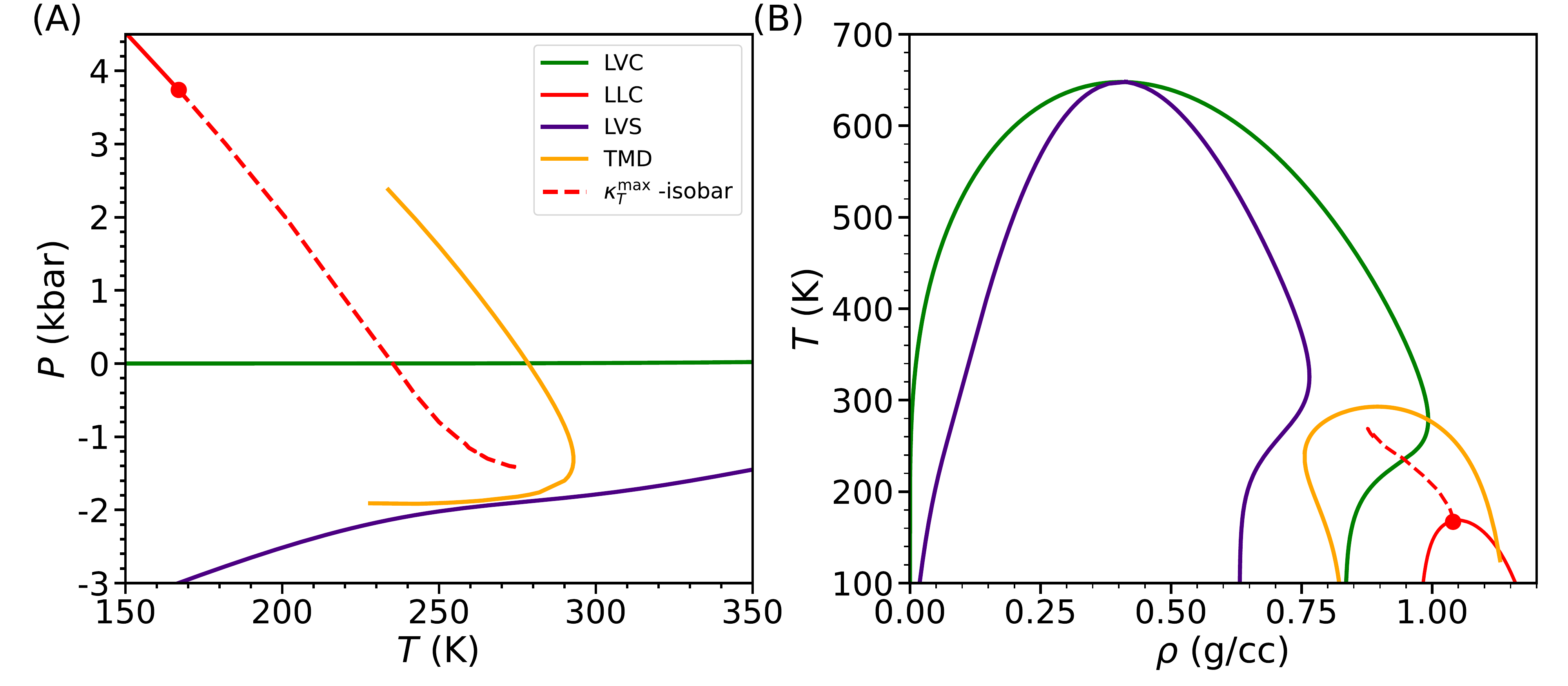}
    \caption{The phase behavior of the model water system showing the liquid-liquid coexistence (LLC), liquid-vapor coexistence (LVC), and the liquid-vapor spinodal (LVS) lines, along with the temperature of maximum density (TMD), the locus of the maximum of isothermal compressibility on isobaric cooling (denoted as $\kappa_T^{\rm max}$-isobar, or the Widom line) in the $P-T$ (A) and $\rho-T$ (B) planes~\cite{pablo_jcp_1999}. The red circle indicates the liquid-liquid critical point (LLCP).} 
    \label{fig_11}
\end{figure}

In Figs.~\ref{fig_11}A and~\ref{fig_11}B, we show the phase diagram of this microscopic model consisting of the liquid-vapor coexistence (LVC) line along with the liquid-liquid coexistence (LLC) line, LLCP (indicated by the red filled circle), the Widom line (defined as the locus of $\kappa_T$ maximum on isobaric cooling and denoted as $\kappa_T^{\max}$-isobar), the TMD line and the LV spinodal (LVS) in the $P-T$ and $\rho-T$ planes~\cite{pablo_jcp_1999}. The coexistence lines can be determined by equating chemical potential and grand potential density of two phases (say, $\alpha$ and $\beta$) at a fixed $T$; $\mu_{\rm fl}^{\alpha}\left(\rho_{\alpha}\right) = \mu_{\rm fl}^{\beta}\left(\rho_{\beta}\right)$, and $\omega_{\rm fl}^{\alpha}(\rho_{\alpha}) = \omega_{\rm fl}^{\beta}(\rho_{\beta})$. Here, $\mu_{\rm fl} = \left(\frac{\partial f_{\rm fl}(\rho)}{\partial\rho}\right)_{T}$, $f_{\rm fl} = F_{\rm fl} / V$ is the Helmholtz free energy density and $\omega_{\rm fl}$ is the grand potential density, $\omega_{\rm fl} = f_{\rm fl} - \mu_{\rm fl} \rho$. The above conditions ensure that the system is in both thermal and mechanical equilibrium at the coexistence condition. 
\subsection{Isothermal Compressibility Anomaly Along Isochores}
\begin{figure} 
    \centering
          \includegraphics[width=\linewidth]{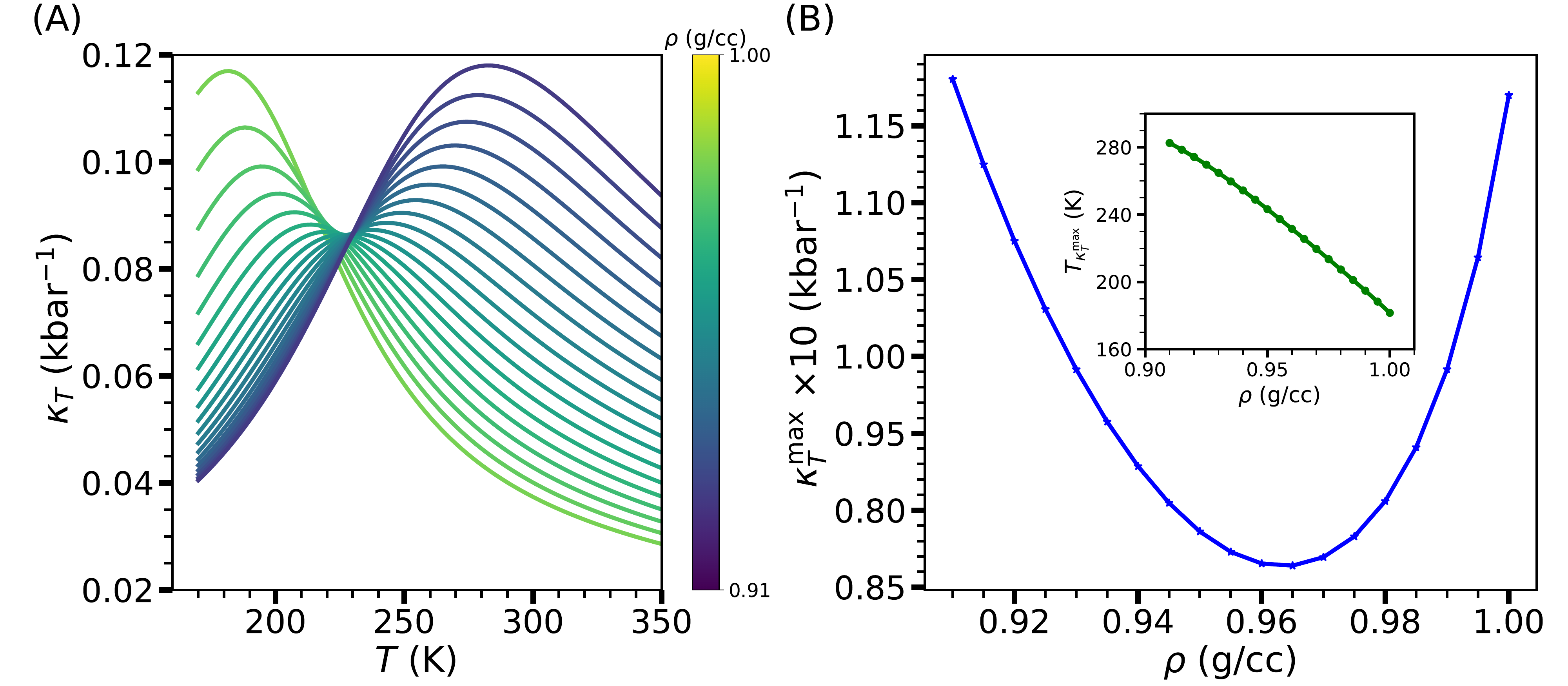}
    \caption{(A) The behavior of the isothermal compressibility $\kappa_T$ on isochoric cooling at different system's densities ($\rho$) is shown. (B) The non-monotonic variation of the maximum of the isothermal compressibility ($\kappa_T^{\rm max}$) with the density $\rho$. In the inset, we have shown the variation of the temperature of maximum $\kappa_T$ ($T_{\kappa_T^{\rm max}}$) on isochoric cooling.} 
    \label{fig_12}
\end{figure}

Recent experimental and computer simulation studies on TIP4P/2005 water reported sound velocity minimum and isothermal compressibility (or density fluctuation) maximum on isochoric cooling at negative pressure conditions~\cite{frederic_pnas_2014,frederic_jpcl_2017, singh_tip4p_2017_2}. These thermodynamic anomalies get pronounced on decompression (or lowering the system's density). In Fig.~\ref{fig_12}A, we report the behavior of isothermal compressibility $\kappa_T$ on isochoric cooling at different densities for the model system studied here. As evident from the figure and also reported in the previous studies~\cite{frederic_pnas_2014,frederic_jpcl_2017, singh_tip4p_2017_2}, $\kappa_T$ shows a non-monotonic dependence on $T$. For the low density isochores, or at negative pressure conditions where the maximum of $\kappa_T$ ($\kappa_T^{\rm max}$) gets pronounced on decompression, the origin of $\kappa_T$ anomaly is attributed to the peculiar shape of water's LV spinodal~\cite{singh_tip4p_2017_2} which is a consequence of the density anomaly~\cite{frederic_jpcl_2017}. The $\kappa_T$ anomaly for the higher density isochores (isochores for which the maximum of $\kappa_T$ gets pronounced on compression) is often attributed to the LLCP~\cite{singh_tip4p_2017_2, frederic_jpcl_2017}. 

In a recent study on TIP4P/2005, Altabet et al.~\cite{singh_tip4p_2017_2} reported that the $T$-dependent $\kappa_T$ for intermediate density isochores ($\rho \sim 0.96$ g/cc) exhibits two maxima. The higher temperature maximum was found to be on the $\kappa_T^{\rm max}$ line emanating from the LV spinodal, and the lower temperature maximum on the $\kappa_T^{\rm max}$ line that emanates from the LLCP. Unlike the TIP4P/2005 water, here we do not observe two separate peaks in $\kappa_T$, rather the strength of the $\kappa_T^{\rm max}$ gradually weakens and then increases with increasing the density (see Fig.~\ref{fig_12}B and also Fig.~\ref{fig_s1} in the Supplementary Materials). This suggests that, for the intermediate density isochores ($\rho \sim 0.96$ g/cc), it is not possible to decouple the effects of the peculiar shape of the LV spinodal and the LLCP on the $\kappa_T^{\rm max}$. For the intermediate density isochores, the system is supercritical with respect to the LLCP and near to the LV spinodal simultaneously, and both of these factors can give rise to the enhanced density fluctuations, or the $\kappa_T^{\rm max}$. Thus, our results suggest that near the ambient pressure, the $\kappa_T$ anomaly of water along isochores can not be solely attributed to the LLCP, and one must also take into account the effects of the peculiar shape of the LV spinodal. In the subsequent sections the isothermal compressibility maximum along isochores is denoted as $\kappa_T^{\rm max}$-isochore, and along isobars as $\kappa_T^{\rm max}$-isobar.    
\section{Vapor Nucleation at Negative Pressures}\label{lv}
\subsection{Liquid-Vapor Surface Tension}\label{lvst}
As the nucleation barrier is quite sensitive to interfacial surface tension, it becomes crucial to study first the behavior of liquid-vapor surface tension ($\gamma_{\rm vl}$) along the coexistence line. In order to calculate $\gamma_{\rm vl}$ using CDFT, first, we need to obtain the equilibrium density profile along the direction perpendicular to the interface (say, $z$-axis) between the coexisting phases ($\rho(z)$). The surface tension is the extra free energy cost for the formation of the interface $\rho(z)$, and is given as, $\gamma_{\rm vl} = \left(\Omega\left(\rho(z)\right) - \Omega_{\rm vap/l}\right)/A$. Here, $\Omega_{\rm vap/l}$ is the grand potential of the coexisting vapor or the liquid phase, and $A$ is the area of the interface. $\Omega \left[\rho (z)\right]$ is the grand canonical free energy functional corresponding to the inhomogeneous system with density profile $\rho(z)$, which within the framework of the square-gradient approximation is given as~\cite{ch_1958, oxtoby_2000}  
\begin{equation} \label{eq:5} 
\Omega[\rho(z)] = \int dz\left[f_{\rm fl}(\rho(z)) - \mu \rho(z)\right] + \frac{1}{2}\int dz 
\left[K_{\rho}\left(\nabla \rho(z)\right)^{2}\right],      
\end{equation} 
where $K_{\rho}$ is related to the correlation length, and $\mu$ is the coexistence chemical potential. In this work, the value of $K_{\rho}$ ($5.0$ in reduced units) is selected in such a way that the calculated surface tension for the microscopic model matches with the reported experimental value at $T = 350$ K~\cite{vega_2007}. The non-local effects in the system due to inhomogeneity in the density are accounted for in the square gradient term. $\rho(z)$ is obtained by solving the Euler-Lagrange equations associated with the following equilibrium condition, $\delta \Omega \left[\rho\left(z\right)\right]/\delta \rho(z) = 0$.   
\begin{figure} [b]
    \centering
          \includegraphics[width=\linewidth]{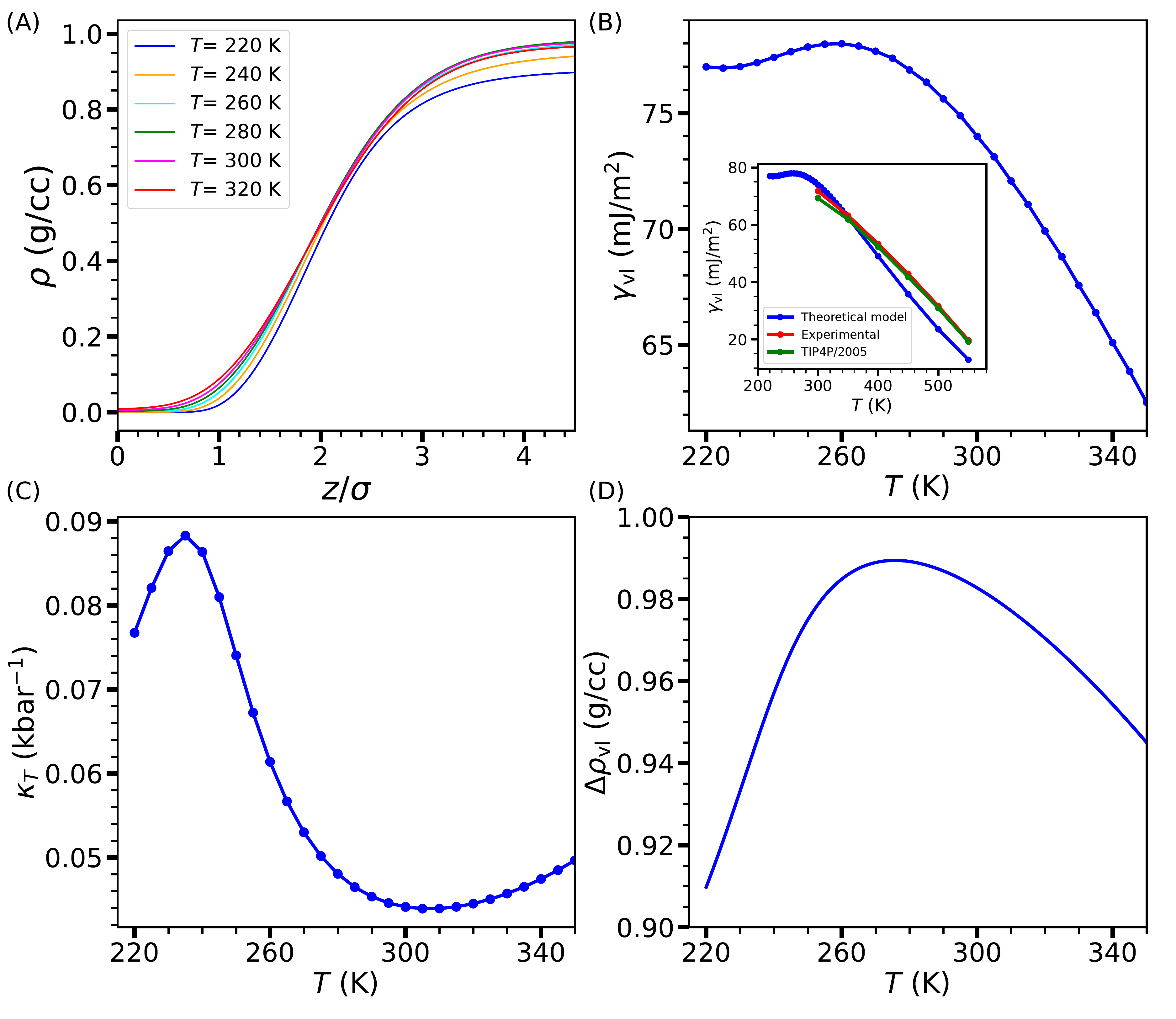}
    \caption{(A) The liquid-vapor density profiles ($\rho(z)$) at different temperatures along the liquid-vapor coexistence line are shown. (B) The calculated liquid-vapor interfacial surface tension ($\gamma_{\rm vl}$) at different temperatures along the coexistence line is shown. In the inset, we have shown the $\gamma_{\rm vl}$ obtained in experiments, computed for the TIP4P/2005 water, and calculated in this work. (C) The $T$-dependent isothermal compressibility ($\kappa_T$) along the liquid-vapor coexistence line is shown. (D) The density difference between the coexisting vapor and liquid phases ($\Delta \rho_{\rm vl}$) at different temperatures is shown.} 
    \label{fig_2}
\end{figure}
In Fig.~\ref{fig_2}A, we show the $T$-dependent density profile $\rho(z)$ across the planar liquid-vapor interface, and in Fig.~\ref{fig_2}B, we report the $T$-dependent $\gamma_{\rm vl}$ calculated using the density profiles shown in Fig.~\ref{fig_2}A. We note a peculiar $T$-dependence of $\gamma_{\rm vl}$ where it increases first with the decrease in $T$ at high temperatures and then decreases and becomes almost insensitive to $T$ in the vicinity of the Widom line ($T \sim 230$ K at the coexistence pressure). One can qualitatively understand this anomalous $T$-dependence of $\gamma_{\rm vl}$ in terms of the $T$-dependence of the density difference between the coexisting phases ($\Delta \rho_{\rm vl}$) and the isothermal compressibility $\kappa_T$ along the coexistence line. The $T$-dependent variation of $\kappa_T$ and $\Delta \rho_{\rm vl}$ is shown in Figs.~\ref{fig_2}C and~\ref{fig_2}D, respectively. The Cahn-Hilliard theory suggests that $\gamma_{\rm vl}$ depends on $\Delta \rho_{\rm vl}$ as, $\gamma_{\rm vl} \sim \Delta \rho_{\rm vl}^2$~\cite{ch_1958, singh_2014}. $\gamma_{\rm vl}$ also depends ``non-trivially" on the isothermal compressibility and decreases with the increases of $\kappa_T$~\cite{kt_gamma_1971, kt_gamma_1974}. This is due to the softening of the free energy surface along the density, which gives rise to a diffused interface, and in turn, the lower surface free energy~\cite{singh_2014}. The initial increase of $\gamma_{\rm vl}$ with the decrease in $T$ could be mainly due to the increase in $\Delta \rho_{\rm vl}$ (see Fig.~\ref{fig_2}D). Below the TMD, the behavior of both $\Delta \rho_{\rm vl}$ and $\kappa_T$ favors a decrease of $\gamma_{\rm vl}$ with the decrease in $T$. In the vicinity and below the Widom line (temperature of maximum $\kappa_T$), the weak $T$-dependence of $\gamma_{\rm vl}$ could be understood in terms of the opposing effects of the $T$-dependence of $\kappa_T$ and $\Delta \rho_{\rm vl}$ on $\gamma_{\rm vl}$. Thus, the observed peculiar $T$-dependence of $\gamma_{\rm vl}$ can be attributed to the anomalous change of $\Delta \rho_{\rm vl}$ and $\kappa_T$ along the coexistence line. This qualitative explanation, however, needs a more careful quantitative validation to unambiguously establish the relative contributions of the behavior of $\kappa_T$ and $\Delta \rho_{\rm vl}$ to $\gamma_{\rm vl}$. We can not also discard the possibility of other ``non-trivial" contributions to the anomalous $T$ dependence of $\gamma_{\rm vl}$.       

We also note that at higher temperatures, the $\gamma_{\rm vl}$ calculated here agrees well with the surface tension reported for the TIP4P/2005 water and in experiments (see the inset of Fig.~\ref{fig_2}B). This validates the choice of our model system and also the choice of the $K_\rho$ value used to study the liquid-vapor interfacial properties. It would be an interesting avenue for future research to validate this observed anomalous $T$-dependence of $\gamma_{\rm vl}$ at metastable (with respect to the ice) conditions using computer simulations on realistic water models, such as TIP4P/2005. The experimental measurement of $\gamma_{\rm vl}$ at lower temperatures (below the freezing temperature) would be challenging due to spontaneous ice crystallization.   
\begin{figure}
    \centering
          \includegraphics[width=\linewidth]{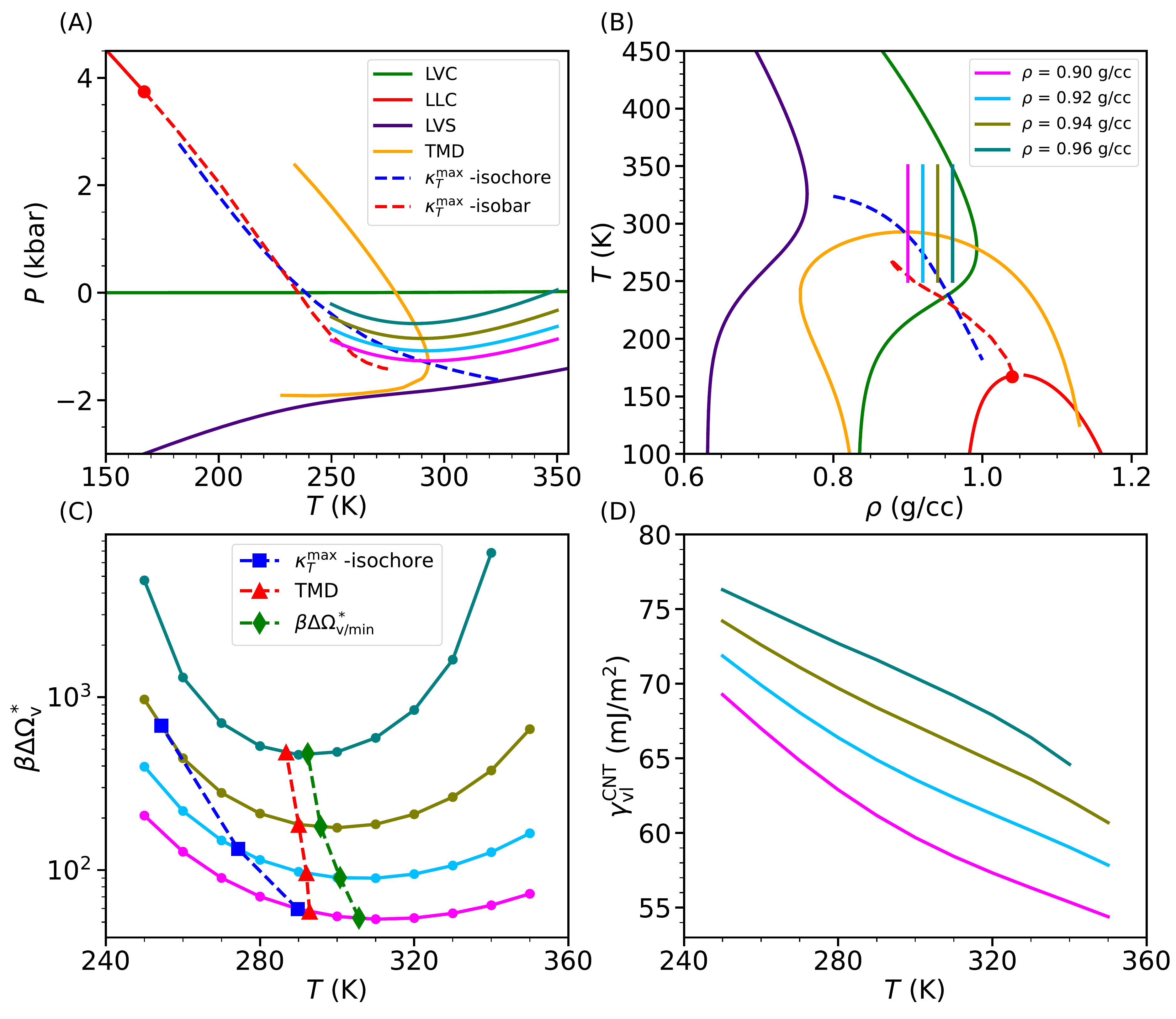}
    \caption{The isochoric cooling pathways in the $P-T$ plane (A) and the $\rho-T$ plane (B) are shown. (C) The $T$-dependent scaled vapor nucleation barrier along different isochores ($\beta \Delta \Omega_{\rm v}^*$) is shown. We note the non-monotonic temperature dependence of $\beta \Delta \Omega_{\rm v}^*$. The red triangles indicate the temperature of maximum density (TMD), the green diamonds indicate the temperature of minimum scaled vapor nucleation barrier on isochoric cooling ($\beta \Delta \Omega^*_{\rm v/min}$), the blue squares indicate the temperature of maximum isothermal compressibility on isochoric cooling ($\kappa_T^{\rm max}$-isochore). (D) The $T$-dependent liquid-vapor surface tension predicted by the CNT ($\gamma_{\rm vl}^{\rm CNT}$) along different isochores is shown.} 
    \label{fig_3}
\end{figure} 
\subsection{Vapor Nucleation on Isochoric Cooling}\label{lvv}
Here, we have studied the vapor nucleation from negative pressure water along different isochoric paths, shown in Figs.~\ref{fig_3}A and 3B in the $P-T$ and $\rho-T$ planes, respectively. On isochoric cooling at negative pressures, the system first crosses the TMD line where the liquid is maximally stretched (metastability with respect to the vapor phase is maximum, see Fig.~\ref{fig_s2} in the Supplementary Materials). On further cooling, the system undergoes enhanced density fluctuations as it crosses the $\kappa_T^{\rm max}$- isochore line. It is reported in the literature that enhanced density fluctuations of the metastable fluid -- a consequence of the flattening of the free energy surface -- facilitate the phase transition by decreasing the nucleation barrier~\cite{frenkel_science, oxtoby_2000}. Thus, both the thermodynamic anomalies, TMD and $\kappa_T^{\rm max}$-isochore, are expected to enhance the vapor nucleation from stretched water. To explore the interplay between these thermodynamic anomalies and the vapor nucleation from the stretched water, in Fig.~\ref{fig_3}C, we have reported the $T$-dependent scaled vapor nucleation barrier ($\beta \Delta \Omega_{\rm v}^*$) along different isochoric paths. $\Delta \Omega_{\rm v}^*$ is calculated using the CDFT~\cite{ch_1958, oxtoby_1996, oxtoby_1998, oxtoby_2000, oxtoby_1994, ssb_2013, singh_2014, bagchi_jcp_2018} which allows us to calculate the free energy of the critical nucleus without making the capillary approximation (unlike, the CNT). In the CDFT, one directly gets the (unstable) equilibrium density profile of the critical cluster by minimizing the grand potential of the inhomogeneous system,
\begin{equation} \label{eq1} 
 \Omega[\rho (\mathbf{r})] = \int d\mathbf{r}\left[f_{\rm fl}(\rho(\mathbf{r})) - \mu\rho(\mathbf{r})\right] +  \frac{1}{2} 
\int d\mathbf{r} \left[K_{\rho } \left(\nabla\rho(\mathbf{r})\right)^{2}\right],      
\end{equation} 
with respect to density profile $\rho (\mathbf{r})$ and solving the resulting ordinary differential equation with boundary conditions, $d\rho/dr = 0$ at $r = 0$ and $\rho = \rho_{\rm l}$ at $r \to \infty$. The nucleation barrier $\Delta \Omega_{\rm v}^*$ is the free energy cost for the formation of the density profile of the critical vapor cluster and is given as $\Delta \Omega_{\rm v}^* = \Omega [\rho (\mathbf{r})] - \Omega(\rho_{\rm l})$. 
\begin{figure}
    \centering
          \includegraphics[width=\linewidth]{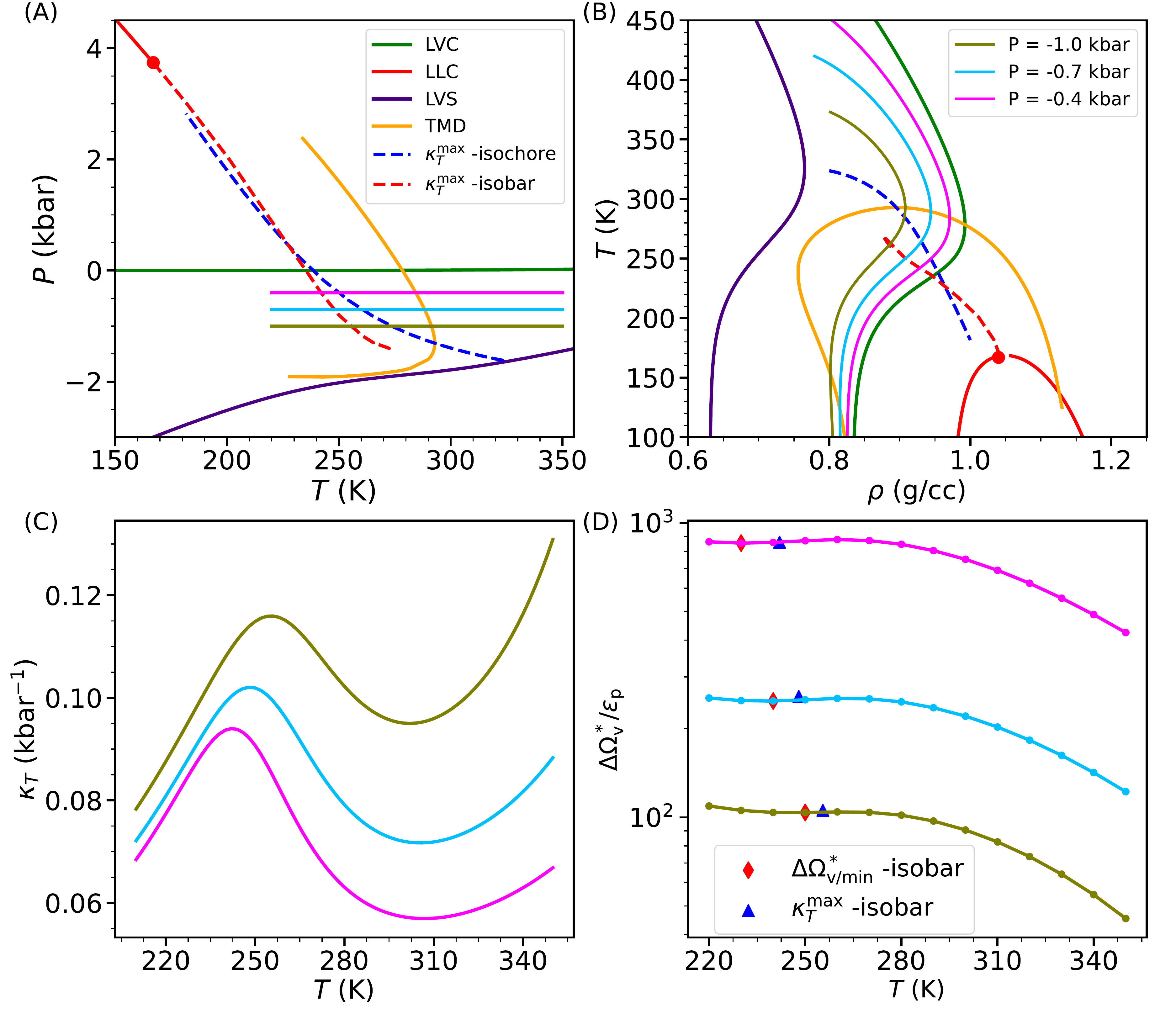}
    \caption{The isobaric cooling pathways in the $P-T$ (A) and the $\rho-T$ (B) planes at negative pressures are shown. (C) The $T$-dependent isothermal compressibility $\kappa_T$ on isobaric cooling at pressures $-400$ bar, $-700$ bar, and $-1000$ bar is shown. (D) The $T$-dependent scaled vapor nucleation barrier ($\Delta \Omega_{\rm v}^* / \epsilon_{\rm p}$) along different isobars is shown. We note that in the close vicinity of the Widom line, $\Delta \Omega_{\rm v}^*$ shows a weak non-monotonic $T$-dependence. The blue triangles indicate the temperature of maximum isothermal compressibility on isobaric cooling ($\kappa_T^{\rm max}$-isobar), and the red diamonds indicate the temperature of minimum vapor nucleation barrier on isobaric cooling in the vicinity of the $\kappa_T^{\rm max}$-isobar (denoted as $\Delta \Omega^*_{\rm v/min}$-isobar).} 
    \label{fig_4}
\end{figure}

We note a non-monotonic $T$-dependence of $\beta \Delta \Omega_{\rm v}^*$ on isochoric cooling (see Fig.~\ref{fig_3}C) where $\beta \Delta \Omega_{\rm v}^*$ decreases with the decrease in temperature, attains a minimum and then increases on further lowering the temperature. We additionally note that the temperature of minimum nucleation barrier ($T_{\beta \Delta \Omega_{\rm v/min}^*}$) does not coincide either with the temperature of maximum metastability (or, the TMD) or with the temperature of maximum density fluctuation (or, $\kappa_T^{\rm max}$-isochore). This observation suggests that the enhanced density fluctuations due to the softening of the metastable liquid free energy basin in the vicinity of the LV spinodal do not lead to any drastic change in the nucleation behavior --- unlike the work by ten Wolde and Frenkel~\cite{frenkel_science} on a model colloidal system with short-range attractive interactions where enhanced density fluctuations (due to the presence of the submerged critical point) were reported to lower the nucleation barrier drastically. This observation further suggests that the minimum in the $\beta \Delta \Omega_{\rm v}^*$ is also not a direct consequence of the maximum metastability that the liquid water attains at the TMD. In a recent experimental study, Caupin and coworkers~\cite{frederic_natphys_2013} studied a single isochore ($\rho = 0.9228$ g/cc) and reported a similar non-monotonic temperature dependence of the nucleation barrier with the $T_{\beta \Delta \Omega_{\rm v/min}^*}$ ($\sim 322$ K) $>$ TMD ($ \sim 296.4$ K). 

These studies suggest that a larger tension in liquid water does not necessarily imply that it is more prone to cavitate. The non-monotonic behavior of $\Delta \Omega_{\rm v}^*$ on isochoric cooling is a consequence of the combined effect of the existence of TMD (where the liquid attains maximum metastability) and the temperature variation of $\gamma_{\rm vl}$. In Fig.~\ref{fig_3}D, we report the temperature variation of the vapor-liquid surface tension calculated using the CNT ($\gamma_{\rm vl}^{\rm CNT}$) along different isochoric cooling pathways; $\Delta \Omega_{\rm v}^* = 16\pi {\gamma_{\rm vl}^{\rm CNT}}^3 / 3(\Delta P)^2$, where $\Delta P$ is the pressure difference between the metastable liquid and the stable vapor phase. We observe a monotonic non-linear increase of $\gamma_{\rm vl}^{\rm CNT}$ on isochoric cooling. Interestingly, however, we also note that the difference between the $T_{\beta \Delta \Omega_{\rm v/min}^*}$ and the TMD decreases on increasing the density of the isochore (see Fig.~\ref{fig_3}C) suggesting a strong correlation between the propensity of cavitation and metastability of the liquid phase. Therefore, one should be cautious about using the onset of cavitation to locate the TMD at negative pressures in fluid inclusion experiments, especially for low-density isochores~\cite{angell_sci}.  
\subsection{Vapor Nucleation on Isobaric Cooling}\label{lvp}
In this section, we have studied the vapor nucleation from stretched and doubly metastable water on isobaric cooling. In Figs.~\ref{fig_4}A and~\ref{fig_4}B, we show the isobaric cooling pathways in the $P-T$ and $\rho-T$ planes, respectively. It is evident from the figure that all the three isobars cross the Widom line in the negative pressure (more precisely, doubly metastable) region of the phase diagram. It would be, therefore, desirable to explore the interplay between the enhanced density fluctuations (or, maximum in the isothermal compressibility on isobaric cooling, $\kappa_T^{\rm max}$-isobar, see Fig.~\ref{fig_4}C) of the (metastable) liquid water and the vapor nucleation barrier in the close vicinity of the Widom line.
 
The $T$-dependent vapor nucleation barrier $\Delta \Omega_{\rm v}^*$ (scaled with $\epsilon_{\rm p}$) for three different isobars are shown in Fig.~\ref{fig_4}D. At higher temperatures, $\Delta \Omega_{\rm v}^*$ increases monotonically on decreasing the temperature. However, at temperatures in the vicinity of the Widom line, $\Delta \Omega_{\rm v}^*$ shows a weak non-monotonic dependence on $T$. The minimum in the nucleation barrier occurs at a temperature very close to the Widom line (see Fig.~\ref{fig_4}D). This observed crossover behavior of $\Delta \Omega_{\rm v}^*$ is an important result because this establishes a direct connection between the thermodynamic (isothermal compressibility) anomaly and the vapor nucleation barrier on isobaric cooling at negative pressures. A similar crossover behavior is also expected for high pressure isobars. It is also worth noting that we did not observe any noticeable effect of the flattening of the free energy surface (or, the isothermal compressibility maximum) on the nucleation barrier along isochores.  

This observed crossover behavior of $\Delta \Omega_{\rm v}^*$ can be explained in terms of the flattening of the metastable liquid's free energy basin near the Widom line (a consequence of the existence of the metastable LLCP, see Fig.~\ref{fig_11}) which gives rise to the enhanced density fluctuations. The free energy flattening reduces the surface tension by making the interfaces more diffused (see Fig.~\ref{fig_2})~\cite{singh_2014}. Therefore, even though the metastability of the vapor phase with respect to the liquid phase decreases monotonically on decreasing the temperature (see Fig.~\ref{fig_s4} in the Supplementary Materials), the decrease of the surface tension in the vicinity of the Widom line cancels out the effects of the decrease in the metastability, and in turn, gives rise to observed crossover behavior. Recent studies of ice nucleation from metastable water report a similar crossover in the $T$-dependence of the (ice) nucleation barrier in the vicinity of the Widom line on isobaric cooling~\cite{poole_2015, singh_2014}. Thus, the metastable LLCP affects both the ice and the vapor nucleation kinetics from the metastable water. Also, this observed non-monotonic dependence of $\Delta \Omega_{\rm v}^*$ on $T$ shows a close resemblance with the enhancement of crystal nucleation near the metastable critical point of a model colloidal fluid with short-range attractive interactions~\cite{frenkel_science, oxtoby_2000}.     
\section{Ice Nucleation from Metastable and Doubly Metastable Water: A Two Order Parameter Description} \label{Is}
\subsection{Ice-Liquid Surface Tension}  \label{lsfes}
The one order parameter description was sufficient to describe the liquid-vapor phase transitions discussed in the previous sections. However, to study the ice-liquid interfacial properties and the nucleation of the ice phase from the metastable liquid water, a two order parameter -- density $\rho$ and structural order parameter $m$ -- description is required. We follow Oxtoby and coworker~\cite{oxtoby_1998, oxtoby_jcp_1995} in describing the free energy functional of the inhomogeneous solid(ice)-liquid system characterized by position-dependent order parameters, $\rho(\mathbf{r})$ and $m(\mathbf{r})$. The proposed free energy (grand potential) functional for this case within the framework of the square-gradient approximation is  
\begin{equation} \label{igp} 
\begin{array}{clc} 
 \Omega[\rho (\mathbf{r}),m(\mathbf{r})] & = \int d\mathbf{r}\left[\omega(\rho(\mathbf{r}),m(\mathbf{r}))\right] \\ 
 & +  \frac{1}{2} 
\int d\mathbf{r} \left[K_{\rho } \left(\nabla\rho(\mathbf{r})\right)^{2} 
   + K_{m}\left(\nabla m(\mathbf{r})\right)^{2}\right],
\end{array} 
\end{equation} 
where $K_{\rho}$ and $K_m$ are related to the correlation lengths for $\rho$ and $m$, respectively. $\omega$ is the grand potential density functional of the average density profile $\rho(\mathbf{r})$ and structural order parameter profile $m(\mathbf{r})$. The square gradient terms account for the nonlocal effects in the system due to inhomogeneity in $\rho$ and $m$.  

For the two order parameter description, we have generalized the fluid phase free energy (grand potential) density ($\omega_{\rm fl}$) by separably introducing structural order parameter contribution to the free energy as 
\begin{equation}
    \omega_{\rm fl}(\rho,m) = \omega_{\rm fl}(\rho) + \frac{1}{2} k_{\rm {fl},m} (m - m_{\rm fl})^2,  
\end{equation}
where $k_{{\rm fl},m}$ is the curvature of the grand potential function along $m$, and $m_{\rm fl}$ is the equilibrium order of the fluid phase. We have chosen $m_{\rm fl} = 0$. The ice phase is incorporated separately through a two order parameter harmonic free energy (grand potential) density function 
\begin{equation} \label{ice_w}
\begin{array}{clc} 
    \omega_{\rm ice}(\rho, m) = \frac{1}{2} k_{\rm {ice},\rho} (\rho - \rho_{\rm ice})^2   & + \frac{1}{2}k_{\rm {ice},m} (m- m_{\rm ice})^2  \\ 
    & + \Delta_{\rm ice}(\mu,T), 
 \end{array}    
\end{equation}
where $k_{\rm {ice},\rho}$ and $k_{\rm {ice},m}$ are the curvatures of the free energy of the ice phase along $\rho$ and $m$, respectively. $\rho_{\rm ice}$ and $m_{\rm ice}$ are the equilibrium density and order, respectively, of the ice phase. $\Delta_{\rm ice}$ is the grand potential density of the equilibrium bulk ice phase at a given thermodynamic condition. Considering the weak ($T, P$) dependence of $\rho_{\rm ice}$ and $m_{\rm ice}$ compared to the values for the liquid water, we have neglected here the ($T, P$) dependence of these quantities. We have chosen $\rho_{\rm ice} = 0.91$ g/cc (equilibrium ice(Ih) density of the TIP4P/Ice water at ambient pressure), $m_{\rm ice} = 0.5$ (chosen to be close to the structural order parameter $Q_6$ value for the ice phase~\cite{pablo_nature_2014}), $k_{\rm {fl}, m} = 12 $, $k_{\rm ice, \rho} =  240$, $k_{\rm ice, m} = 240$, $K_{\rho} = 5.0$, and $K_{m} = K_{\rho}/2$ (in reduced units) in our calculations. The values of these parameters are chosen in such a way that the ice(Ih)-liquid surface tension at $1$ bar comes out to be close to the experimentally reported surface tension value ($29.6$ $\rm mJ/m^2$)~\cite{gamma_ice_1992, vega_ice_gamma_2016}. We note that Ice(Ih) is the most stable ice polymorph at the ($T, P$) conditions studied in this work~\cite{huang_sci_2016}. The grand potential of the system $\omega$ (see Eq.~\ref{igp}) is given as $\omega = \min(\omega_{\rm fl}, \omega_{\rm ice})$.

\begin{figure}
    \centering
     \includegraphics[width=0.8\linewidth]{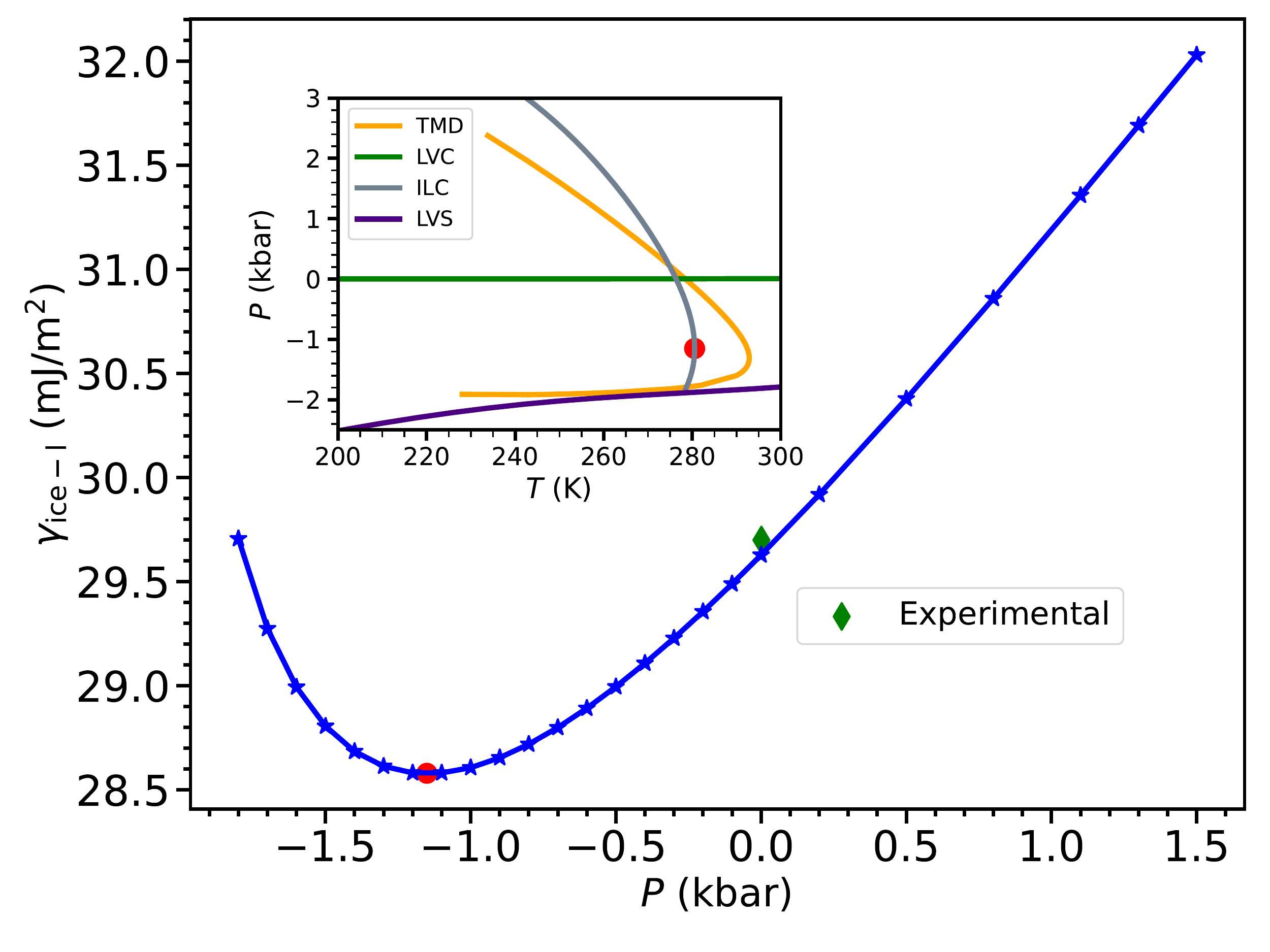}
    \caption{The pressure ($P$) dependence of the ice-liquid surface tension ($\gamma_{\rm ice-l}$) along the coexistence line is shown. We note the non-monotonic dependence of $\gamma_{\rm ice-l}$ on $P$. The green diamond shows the experimentally reported $\gamma_{\rm ice-l}$ at ambient pressure~\cite{gamma_ice_1992}. In the inset, we have shown the ice(Ih)-liquid coexistence line for the TIP4P/Ice water~\cite{vega_prl_2021} along with the TMD, liquid-vapor coexistence (LVC) and the liquid-vapor spinodal (LVS) lines for the water model studied here in the $P-T$ plane. The ice(Ih)-liquid coexistence line for TIP4P/Ice water is shifted vertically in such a way that $\rho_{\rm l} = 0.91$ g/cc (density of the ice-Ih phase near ambient pressure for TIP4P/Ice) at the temperature where $dT/dP = 0$ (indicated by the filled red circle) along the coexistence line.}
    \label{fes}
\end{figure}

Considering the distinct lack of information about the experimental ice(Ih)-liquid coexistence line at negative pressures, here we have used the coexistence line reported for the TIP4P/Ice water to mimic the ice-liquid coexistence line for our model system. This coexistence line is shifted vertically in such a way that $\rho_{\rm l} = 0.91$ g/cc (density of the ice(Ih) phase near ambient pressure for TIP4P/Ice) at the temperature where $dT/dP = 0$ (indicated by the filled red circle in the inset of Fig.~\ref{fes}) along the coexistence line. The coexistence pressure where $dT/dP = 0$ is denoted as $P^* $. We also note that, $\Delta_{\rm ice} = \omega_{\rm fl}$ along the (ice-liquid) coexistence line. The ice-liquid surface tension ($\gamma_{\rm ice-l}$) is the extra free energy cost for the formation of equilibrium density and (structural) order profiles and is given as $\gamma_{\rm ice-l} = \left(\Omega\left(\rho(z),m(z)\right) - \Omega_{\rm ice/l}\right)/A$, where $\Omega_{\rm ice/l}$ is the grand potential of the coexisting ice or the liquid phase, $\rho(z)$ and $m(z)$ are the equilibrium density and order profiles, respectively. $\rho(z)$ and $m(z)$ are obtained by solving the following Euler-Lagrange equations under appropriate 
boundary conditions ($\rho(z) = \rho_{\rm l}^{\rm coex}$ and $m(z) = 0$ at $z = 0$; and $\rho(z) = \rho_{\rm ice}$ ($0.91$ g/cc) and $m(z) = 0.5$ at $z \to \infty$), 
\begin{equation}
\frac{\delta \Omega}{\delta \rho(z)} = 0; \frac{\delta \Omega}{\delta m(z)} = 0, 
\end{equation}
where $\Omega$ is given by Eq.~\ref{igp}. 

In Fig.~\ref{fes}, we report the computed $\gamma_{\rm ice-l}$ along the ice-liquid coexistence line. We found that $\gamma_{\rm ice-l}$ decreases with a decrease in $P$, attains a minimum at $P^*$ ($-1.15$ kbar), and then increases on further lowering the pressure. This anomalous (non-monotonic) dependence of $\gamma_{\rm ice-l}$ on $P$ is a consequence of the non-monotonic change in the absolute value of the density difference between the coexisting phases ($|\Delta \rho_{\rm ice-l}| = |\rho_{\rm ice}- \rho_{\rm l}|$), which in turn also gives rise to the retracing behavior of the coexistence line in the $P-T$ plane (see the inset of Fig.~\ref{fes}). The density difference between the coexisting phases decreases with the decrease in pressure of the system and is zero at the pressure where $dT/dP = 0$. Below $P^*$, liquid water has a lower density than the ice phase. Thus, in the close vicinity of $P^*$, the contribution of the density profile to $\gamma_{\rm ice-l}$ is negligibly small (zero at $P^*$), and hence, order profile is the only contributor to the surface tension. This gives rise to the minimum  $\gamma_{\rm ice-l}$ value at $P^*$. The density profile contribution to $\gamma_{\rm ice-l}$ increases on further lowering the pressure below $P^*$. Hence, non-monotonic dependence of $\gamma_{\rm ice-l}$ on $P$ is predominantly guided by the $P$-dependent change in the contribution of the density profile to the $\gamma_{\rm ice-l}$ along the ice-liquid coexistence line. We shall discuss the consequences of this anomalous $P$ dependence of $\gamma_{\rm ice-l}$ and the retracing behavior of the ice-liquid coexistence line on the ice nucleation in the next section. 
\begin{figure} 
    \centering
          \includegraphics[width=0.8\linewidth]{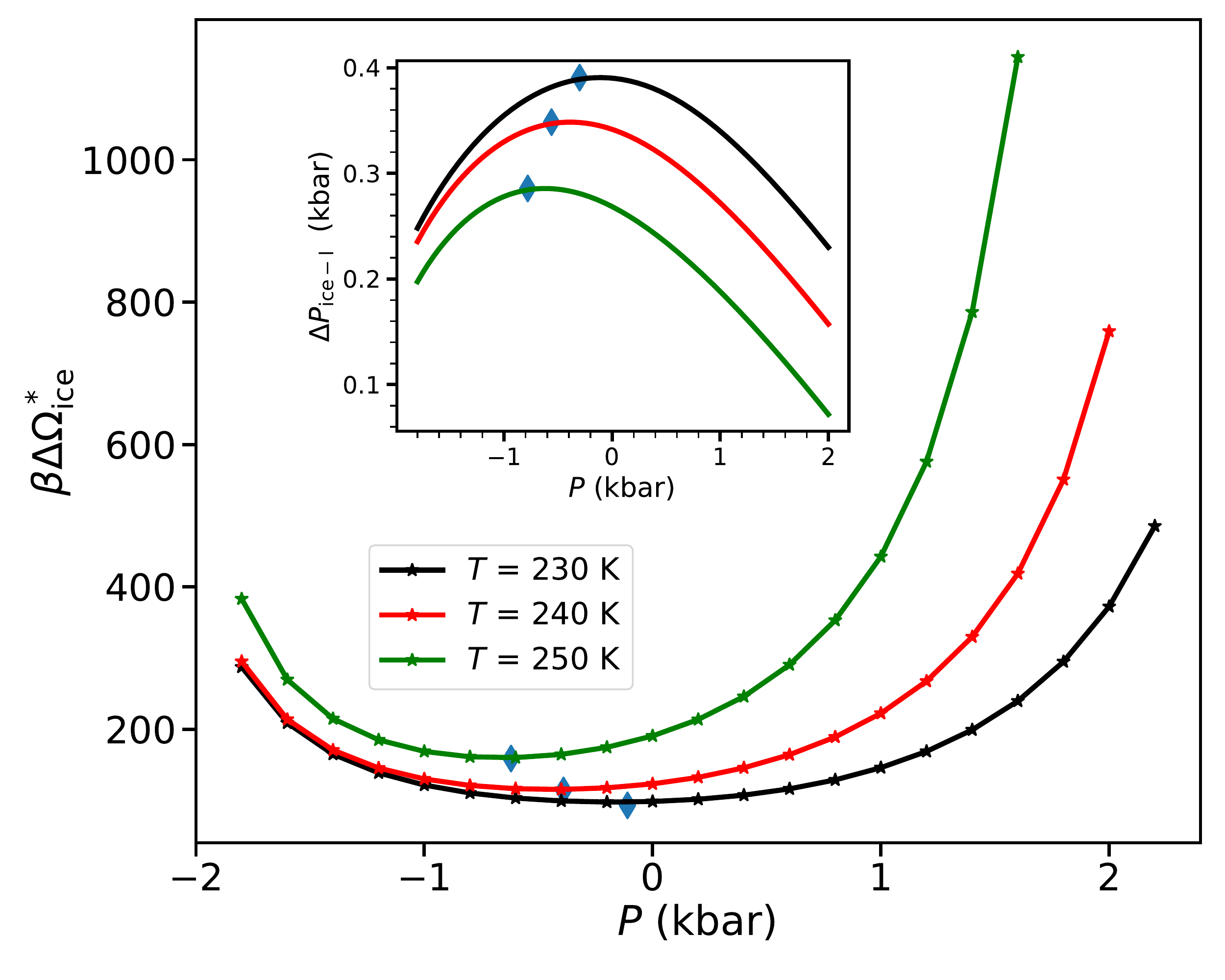}
    \caption{The scaled ice nucleation barrier ($\beta \Delta \Omega_{\rm ice}^*$) vs. pressure of the metastable liquid water ($P$) along three different isotherms is shown. Note the non-monotonic dependence of $\beta \Delta \Omega_{\rm ice}^*$ on $P$ ($\beta$ is constant along an isotherm). The minimum of the  $\beta \Delta \Omega_{\rm ice}^*$ vs. $P$ lies at a pressure where the liquid density is the same as the ice density ($\Delta \rho_{\rm ice-l} = 0$) along an isotherm (marked with filled diamonds). In the inset, we have shown the pressure difference between the (metastable) liquid and the ice phase ($\Delta P_{\rm ice-l}$) at a given chemical potential as a function of $P$. We note a non-monotonic dependence of $\Delta P_{\rm ice-l}$ on $P$ with the position of maximum roughly coinciding with $\Delta \rho_{\rm ice-l} = 0$ (marked with filled diamonds).} 
    \label{fig_6}
\end{figure} 
\subsection{Ice Nucleation Along Isotherms}\label{lst}
As discussed in Section~\ref{intro}, Vega and coworkers~\cite{vega_prl_2021} reported recently an anomalous retracing behavior of the ice nucleation barrier along isotherms where nucleation barrier reaches a minimum at negative pressures in the doubly metastable region. They attributed this anomaly in the nucleation barrier (or, rate) to the reentrance of the ice(Ih)-liquid coexistence line at negative pressures. Here, we focus on understanding this observed interplay between the reentrant behavior of the ice(Ih)-liquid coexistence line and the ice nucleation barrier at negative pressures within the framework of the theoretical model studied in this work. 

The grand potential density of the equilibrium ice phase $\Delta_{\rm ice}$ (see Eq.~\ref{ice_w}) at different metastable state points along an isotherm was estimated by using thermodynamic integration, 
\begin{equation}\label{eq_eos_1}
    \Delta \mu = \mu -\mu_{\rm ice/l}^{\rm coex} =  \int_{P_{\rm ice}}^{P_{\rm coex}} v_{\rm ice}(P) dP,
\end{equation}
where $v_{\rm ice}$ is the specific volume of the ice phase, $\mu$ is the chemical potential of the liquid water, and $\mu_{\rm ice/l}^{\rm coex}$ is the coexistence (ice and liquid) chemical potential. $P_{\rm coex}$ is the coexistence ice pressure, and $P_{\rm ice}$ (note, $-P_{\rm ice} = \Delta_{\rm ice}$) is the ice pressure when liquid water is metastable with respect to the ice phase. The specific volume of the ice was kept constant at $v_{\rm ice}(P) = v_{\rm ice}^0 = 1 / \rho_{\rm ice}^0$ ($\rho_{\rm ice}^0 = 0.91$ g/cc). The ice nucleation barrier ($\Delta \Omega_{\rm ice}^*$) is the free energy cost for the formation of the (unstable) equilibrium density and order profiles (for the two order parameter case) of the critical ice cluster (see Section~\ref{lvv} for the details), and is given as $\Delta \Omega_{\rm ice}^* = \Omega[\rho (\mathbf{r}),m(\mathbf{r})] - \Omega(\rho_{\rm l}, m_{\rm l})$. The density and order profiles can be obtained from the solutions of the Euler-Lagrange equations associated with the following conditions,
\begin{equation}
\frac{\delta \Omega}{\delta \rho(\mathbf r)} = 0; \frac{\delta \Omega}{\delta m(\mathbf r)} = 0. 
\end{equation} 
The ice nucleation barrier along three different isotherms -- $230$ K, $240$ K, and $250$ K --  is shown in Fig.~\ref{fig_6}. Similar to the observations made by Vega and coworkers~\cite{vega_prl_2021}, we note a non-monotonic $P$ dependence of $\Delta \Omega_{\rm ice}^*$ along isothermal paths where $\Delta \Omega_{\rm ice}^*$ decreases initially and then increases on decreasing the pressure. To gain deeper insights into the origin of this non-monotonic (or, retracing) behavior of $\Delta \Omega_{\rm ice}^*$, in the inset figure, we report the $P$-dependent pressure difference between the metastable liquid water and the stable ice phase $\Delta P_{\rm ice-l} (\mu, T) = P_{\rm l} (\mu, T) - P_{\rm ice} (\mu, T)$ along different isotherms studied here. The computed $\Delta P_{\rm ice-l}$ also shows a non-monotonic $P$-dependence along isothermal paths. The maximum of the $\Delta P_{\rm ice-l}$ lies close to the pressure where liquid water density is equal to the ice density along the corresponding isotherm ($\Delta \rho_{\rm ice-l} = 0$ pressure along isotherms is marked with the filled diamonds). The liquid water expands upon freezing above the pressure of the minimum ice nucleation barrier along an isotherm (denoted as $P_{\Delta \Omega_{\rm ice/min}^*}$) and contracts below $P_{\Delta \Omega_{\rm ice/min}^*}$ (see also Fig.~\ref{fig_s5} in the Supplementary Materials). We also note that the $P_{\Delta \Omega_{\rm ice/min}^*}$ roughly coincides with the pressure of maximum $\Delta P_{\rm ice-l}$. 

It is worth noting that the behavior of $\Delta \Omega_{\rm ice}^*$ is not solely guided by the pressure dependence of $\Delta P_{\rm ice-l}$ but also on the $\gamma_{\rm ice-l}$. As shown in the previous section, $\gamma_{\rm ice-l}$ along the coexistence line shows a minimum at the pressure where $\Delta \rho_{\rm ice-l} = 0$. Following the same argument, it is expected that the surface tension between the ice and metastable liquid ($\gamma_{\rm ice-ml}$) would also show a minimum at the pressure where $\Delta \rho_{\rm ice-l} = 0$ as order profile would again be the only contributor to the surface tension (the reported non-monotonic dependence of the surface tension for TIP4P/Ice at metastable conditions~\cite{vega_prl_2021} supports this argument). By invoking the CNT, the anomalous non-monotonic behavior of $\Delta \Omega_{\rm ice}^*$ can be explained in terms of the non-monotonic behavior of $\Delta P_{\rm ice-l}$ and $\gamma_{\rm ice-ml}$, as the study of Vega and coworkers~\cite{vega_prl_2021} also suggests. Thus, within the framework of our generalized phenomenological model, we are able to capture the anomalous retracing behavior of $\Delta \Omega_{\rm ice}^*$ along isothermal paths at negative pressures. We have not calculated the behavior of the ice nucleation barrier along isobaric paths in the negative pressure region as this dependence can be inferred from Fig.~\ref{fig_6} itself.  
\section{Phase Diagram}\label{phase_diag}
\begin{figure}
    \centering
          \includegraphics[width=0.9\linewidth]{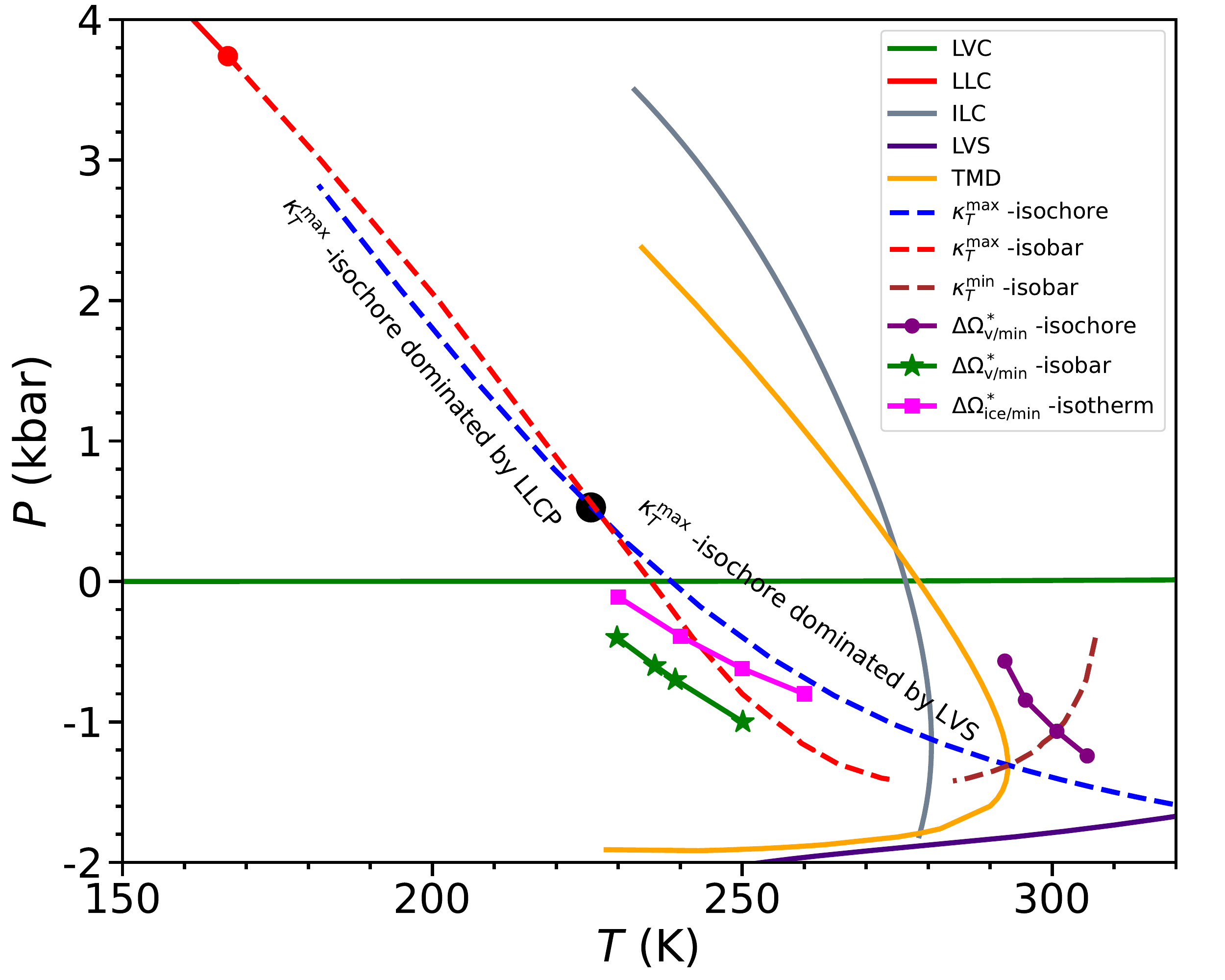}
    \caption{Phase diagram of the phenomenological model summarizing anomalous nucleation kinetics in the $P-T$ plane. Here, we report the locus of minimum free energy barrier for vapor nucleation along isochoric and isobaric cooling paths (indicated by $\Delta \Omega_{\rm v/min}^*$ - isochore, and $\Delta \Omega_{\rm v/min}^*$ - isobar, respectively), along with the locus of the minimum ice nucleation barrier along isotherms ($\Delta \Omega_{\rm ice/min}^*$ - isotherm) in the negative pressure regime of the phase plane. We note that the $\kappa_T^{\rm max}$-isochore line crosses the $\kappa_T^{\rm max}$-isobar line in the $P-T$ plane at the temperature where $\kappa_T^{\rm max}$-isochore shows a minimum on decompression (see Fig.~\ref{fig_12}B). The black filled circle separates the LLCP dominated $\kappa_T^{\rm max}$-isochore and the LV spinodal dominated $\kappa_T^{\rm max}$-isochore (see Figs.~\ref{fig_12}A and~\ref{fig_12}B).} 
    \label{fig_7}
\end{figure}

We have summarized the anomalous nucleation of the vapor and ice phases from metastable liquid water in the phase diagram shown in Fig.~\ref{fig_7}. Here, we have reported the locus of minimum free energy barrier for vapor nucleation along isobaric and isochoric cooling paths ($\Delta \Omega_{\rm v/min}^*$ - isobar and $\Delta \Omega_{\rm v/min}^*$ - isochore, respectively, see Figs.~\ref{fig_3}C and~\ref{fig_4}D), along with the locus of the minimum ice nucleation barrier along isotherms, $\Delta \Omega_{\rm ice/min}^*$ - isotherm (see Fig.~\ref{fig_6}) in the negative pressure regime. Furthermore, we have also reported the $\kappa_T^{\rm max}$ along isochores ranging from low to high densities. Interestingly, we note that the $\kappa_T^{\rm max}$-isochore line crosses the $\kappa_T^{\rm max}$-isobar line in the $P-T$ plane at the temperature where $\kappa_T^{\rm max}$-isochore shows a minimum on decompression (or, lowering the density of liquid water, see Fig.~\ref{fig_12}B). This result suggests that the peculiar shape of the LV spinodal is responsible for the distortion of the $\kappa_T^{\rm max}$-isochore line at moderately high and negative pressures in the $P-T$ plane. Above the crossing temperature, the LLCP contributes dominantly to the $\kappa_T^{\rm max}$-isochore, and below this temperature (for low density isochores), the peculiar shape of the LV spinodal contributes dominantly to the $\kappa_T^{\rm max}$-isochore (see also Fig.~\ref{fig_12}B). Therefore, one must consider the effects of the LLCP as well as the LV spinodal while understanding the origin of the anomalous behavior of thermodynamic response functions (such as sound velocity, isothermal compressibility, etc.) on isochoric cooling, especially at moderately high and negative pressure conditions~\cite{frederic_pnas_2014}.      
\section {Conclusions}\label{conclusions}
In this work, we have employed CDFT to explore the interplay between the thermodynamic anomalies and nucleation (vapor and liquid) kinetics from metastable water, especially at negative pressures. As there is a distinct lack of free energy functional which can simultaneously describe the phase behavior of the liquid water and the ice, in this work we had to constrain to a phenomenological (not microscopic) free energy functional for the ice phase. 

For the model water studied here, we found that, unlike the TIP4P/2005 water, it is not possible to decouple the effects of the peculiar shape of the LV spinodal and the LLCP on the anomalous behavior of $\kappa_T$ along isochores, especially for the intermediate density isochores ($\rho \sim 0.96$ g/cc). Therefore, to precisely understand the origin of the enhanced density fluctuations along isochores near ambient ($T, P$) conditions, one must also take into account the effects of the peculiar shape of the LV spinodal, along with the effects of the (hypothesized) LLCP.  We have further explored the interfacial surface tension and nucleation of the vapor and ice phases from the metastable water. Our results suggest a peculiar $T$-dependence of $\gamma_{\rm vl}$ where it increases first with the decrease in $T$ and then decreases and becomes almost insensitive to $T$ in the vicinity of the Widom line on further lowering $T$. This $T$-dependence can be attributed to the $T$-dependence of the density difference between the coexisting phases $\Delta \rho_{\rm vl}$ and the isothermal compressibility $\kappa_T$ along the coexistence line. The vapor nucleation barrier $\Delta \Omega_{\rm v}^*$ shows a non-monotonic $T$-dependence on isochoric cooling. We also note that the minimum vapor nucleation barrier temperature $T_{\Delta \Omega_{\rm v/min}^*}$ coincides neither with the temperature of maximum $\kappa_T$ nor with the TMD where metastability is maximum. Interestingly, however, we also note that the temperature difference between the $T_{\Delta \Omega_{\rm v/min}^*}$ and the TMD decreases on increasing the density of the isochore, suggesting a stronger correlation between the propensity of cavitation and metastability of the liquid water for higher density isochores. Therefore, one should be cautious about using the onset of cavitation to locate the TMD at negative pressures in fluid inclusion experiments, especially for low-density isochores~\cite{angell_sci} as the minimum in the $\Delta \Omega_{\rm v}^*$ is not a direct consequence of the maximum metastability that the liquid attains at TMD but is a consequence of the combined effect of the TMD and the temperature variation of $\gamma_{\rm vl}$. 

The vapor nucleation barrier $\Delta \Omega_{\rm v}^*$ along isobars also shows a crossover behavior in its $T$-dependence near the Widom line on isobaric cooling. At higher temperatures, $\Delta \Omega_{\rm v}^*$ increases monotonically on decreasing the temperature. However, at temperatures in the vicinity of the Widom line, we observe that $\Delta \Omega_{\rm v}^*$ shows a non-monotonic dependence on $T$. The computational validation of the anomalous vapor nucleation along isobaric and isochoric paths from metastable water at negative pressures would be an interesting avenue for future research. Furthermore, we have also explored the ice nucleation from metastable and double metastable water. Our results on the ice nucleation validate theoretically the anomalous retracing barrier of the ice nucleation barrier along isotherms reported recently in a computational study on TIP4P/Ice water by Vega and coworkers~\cite{vega_prl_2021} and further confirms that the reentrant ice(Ih)-liquid coexistence line can induce a drastic change in the kinetics of ice nucleation. 

Finally, apart from theoretical curiosity, this study provides deeper insights into the nature of the phase transitions (ice and vapor) in metastable water and their interplay with the observed thermodynamic anomalies (such as TMD, isothermal compressibility maximum on isobaric and isochoric cooling, and the retracing behavior of the ice-liquid coexistence) at negative pressure conditions. Recent studies suggest that colloidal tetrahedral liquids~\cite{sciortino_natphys_2014, sciortono_jcp_2019} and other tetrahedral network forming liquids~\cite{sastry_nat_2003, sastry_natphys_2011, sastry_jcp_2021, chen_2017, palmer_pccp_2018, palmer_chem_rev_2018} also show water-like anomalies. Hence, in principle, one can design a tetrahedral network forming (colloidal) fluid where one can selectively enhance the crystallization or vaporization by controlling the thermodynamic conditions. On a more general note, this study establishes a direct connection between the nature of the underlying free energy surface and its interplay with the nucleation kinetics, and this has applicability beyond the tetrahedral network forming liquids.

\begin{acknowledgments}
R.S.S. gratefully acknowledges financial support from DST-SERB (Grant No. SRG/2020/001415) and Indian Institute of Science Education and Research (IISER) Tirupati. Y. S. acknowledges financial support from IISER Tirupati. M. S. acknowledges financial support from DST-SERB (Grant No. SRG/2020/001385). 
\end{acknowledgments}


\bibliography{water}

\pagebreak
\widetext
\begin{center}
\textbf{\large Supplementary Materials}
\end{center}
\setcounter{equation}{0}
\setcounter{figure}{0}
\setcounter{table}{0}
\setcounter{page}{1}
\setcounter{section}{0}
\makeatletter
\renewcommand{\theequation}{S\arabic{equation}}
\renewcommand{\thefigure}{S\arabic{figure}}
\renewcommand{\bibnumfmt}[1]{[S#1]}
\renewcommand{\citenumfont}[1]{S#1}
\begin{figure}[htbp]
    \centering
          \includegraphics[width=0.4\linewidth]{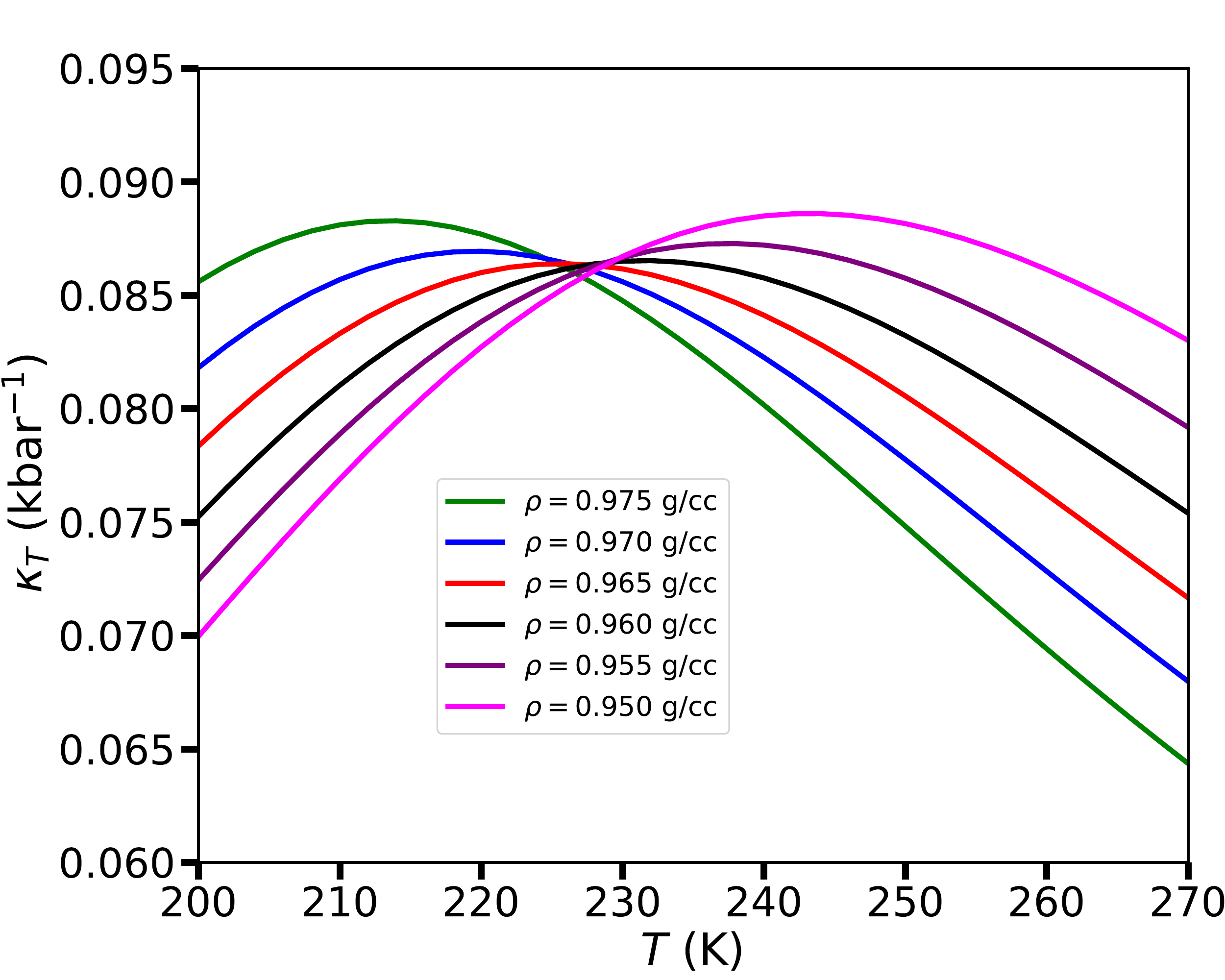}
    \caption{The $T$-dependent isothermal compressibility $\kappa_T$ of liquid water along different isochores is shown. Note that the $\kappa_T$ maximum (denoted as $\kappa_T^{\rm max}$) shows a non-monotonic behavior where it first decreases and then increases on compression (decreasing the density of the system). This crossover behavior suggests that, for low density isochores, the dominant contribution to $\kappa_T^{\rm max}$ comes from the peculiar shape of the LV spinodal, and for high density isochores the LLCP contributes dominantly to the $\kappa_T^{\rm max}$.} 
    \label{fig_s1}
\end{figure} 
\begin{figure}[htbp]
    \centering
          \includegraphics[width=0.4\linewidth]{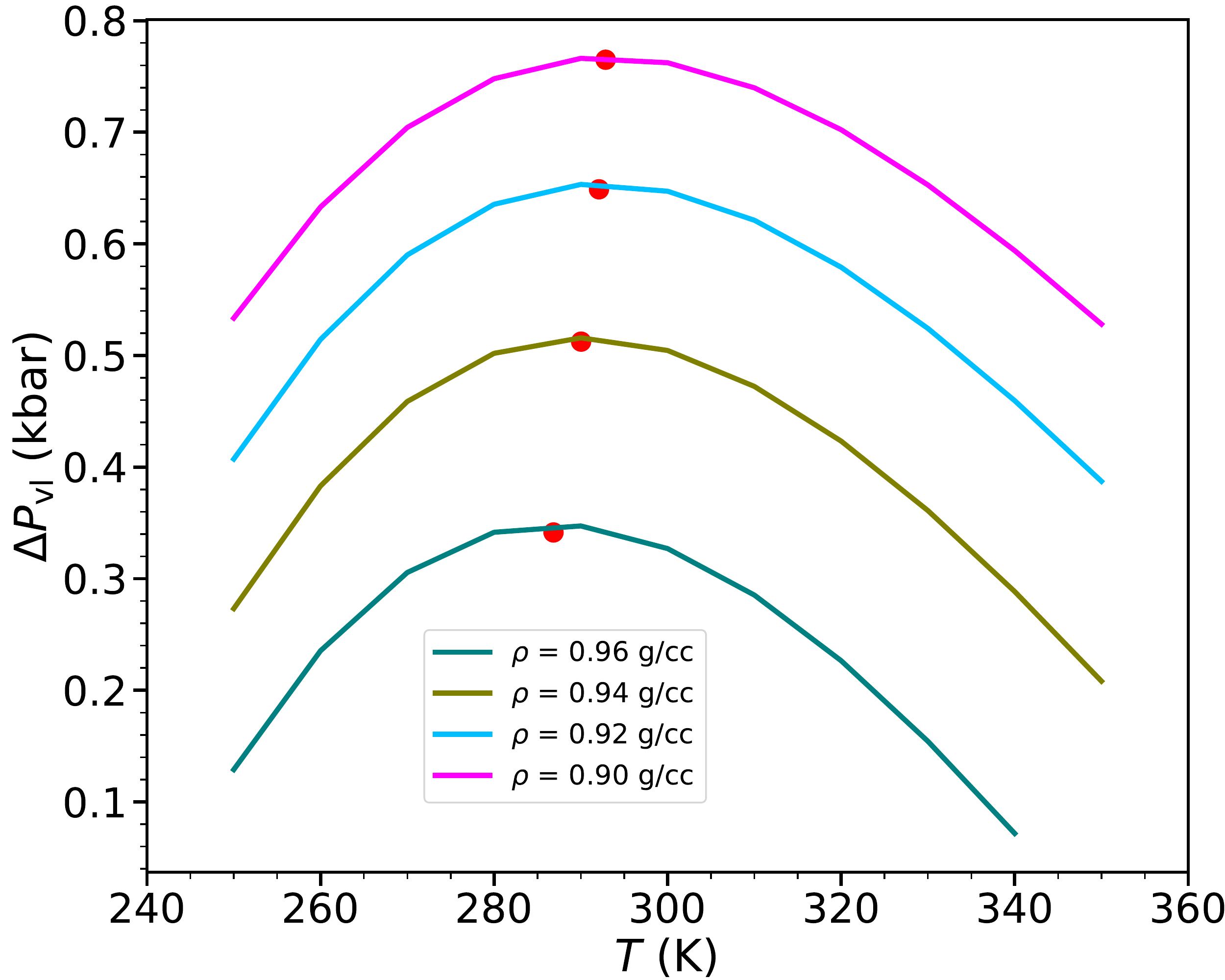}
    \caption{The $T$-dependent pressure difference between the metastable liquid water and the stable vapor phase ($\Delta P_{\rm vl}$) on isochoric cooling the liquid water at different densities is reported. The red filled circles indicate the TMD. We note that the liquid water is maximally stretched at the TMD.} 
    \label{fig_s2}
\end{figure} 
\begin{figure}[htbp]
    \centering
          \includegraphics[width=0.4\linewidth]{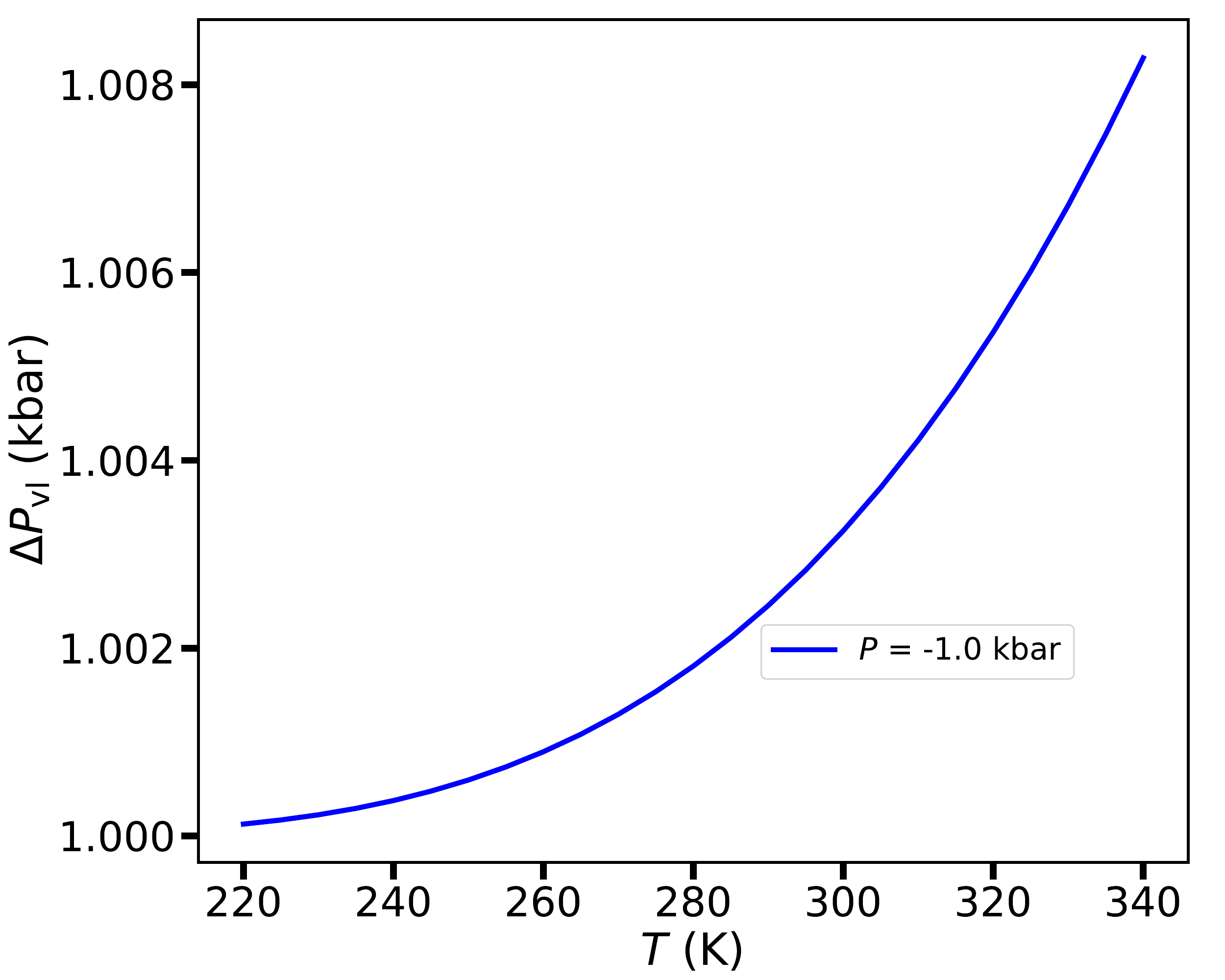}
    \caption{The $T$-dependent pressure difference between the metastable liquid and the stable vapor phase ($\Delta P_{\rm vl}$) on isobaric cooling the liquid water at $-1.0$ kbar is reported.} 
    \label{fig_s4}
\end{figure} 
\begin{figure}[htbp]
    \centering
          \includegraphics[width=0.4\linewidth]{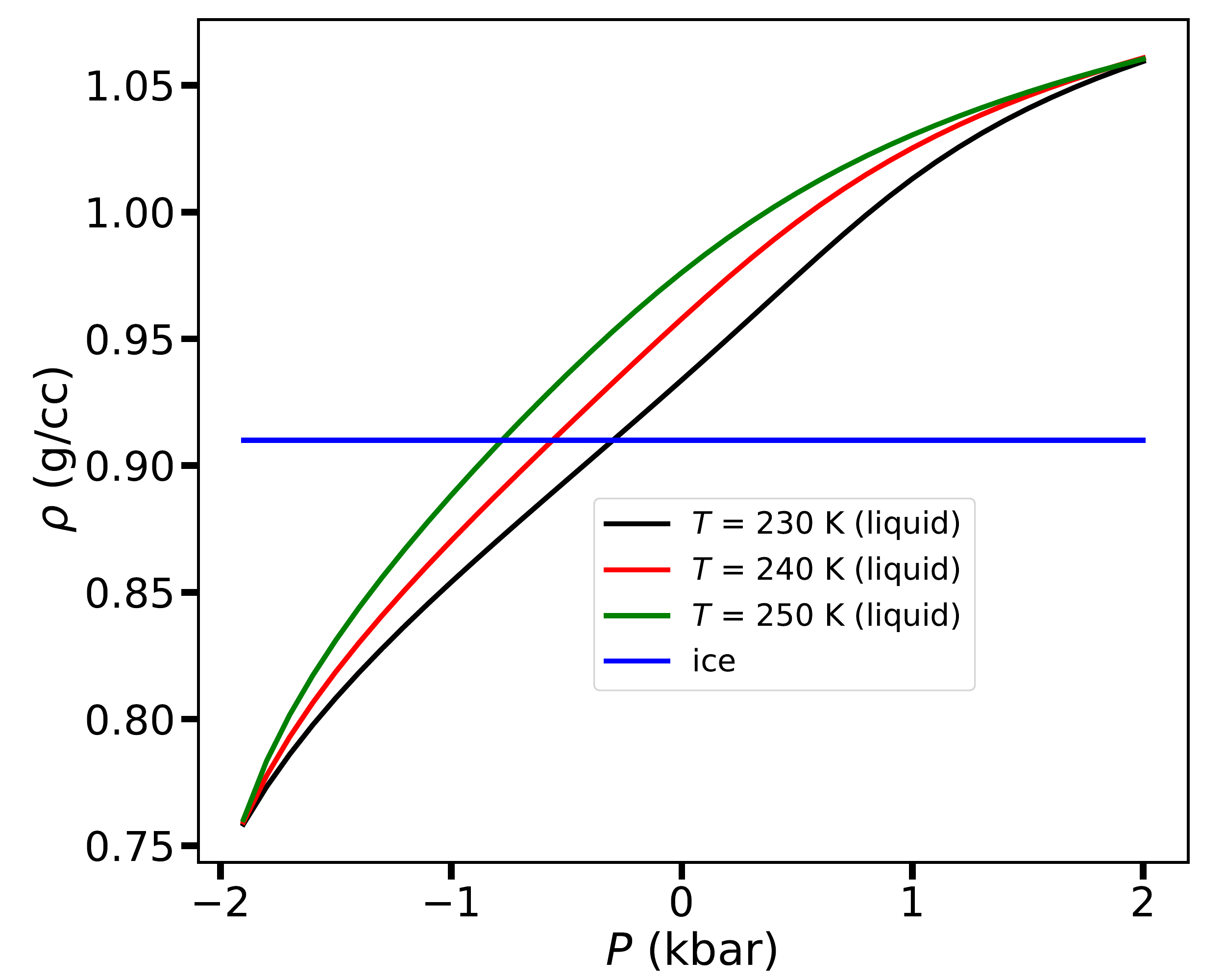}
    \caption{The density of the liquid water along different isotherms on varying the pressure ($P$). The ice density is assumed to be independent of the thermodynamic condition in the ($T, P$) range studied in this work.} 
    \label{fig_s5}
\end{figure}

\end{document}